\documentclass[twocolumn,aps,showpacs]{revtex4}

\usepackage[T1]{fontenc}
\usepackage[latin1]{inputenc}
\usepackage{subfigure}
\usepackage{graphicx}% Include figure files
\usepackage{amssymb}
\usepackage{dcolumn}% Align table columns on decimal point
\usepackage{bm}% bold math

\renewcommand{\d}{\mathrm{d}}
\newcommand{\rhoA}{\hat{\rho}^{(A)}}
\newcommand{\ent}{\mathcal{E}}
\newcommand{\ket}[1]{|#1\rangle}

\begin{document}

\title{Configuration-Space Location of the Entanglement between Two Subsystems}

\author{H.-C. Lin}\email{ho.lin@ucl.ac.uk}
\author{A.J. Fisher}\email{andrew.fisher@ucl.ac.uk}

\affiliation{UCL Department of Physics and Astronomy and London Centre for Nanotechnology,\\ 
University College London, Gower Street, London WC1E 6BT, U.K.}

\begin{abstract}
In this paper we address the question: where in configuration space
is the entanglement between two particles located?  We present a thought-experiment, equally applicable to discrete or continuous-variable systems, in which one or both  parties makes a preliminary measurement of the state with only enough resolution to determine whether or not the
particle resides in a chosen region, before attempting to make use of the entanglement.  We argue that this provides an operational answer to the question of how much entanglement was originally located within the chosen region.  We illustrate the approach in a spin system, and also in a pair of coupled harmonic oscillators.
Our approach is particularly
simple to implement for pure states, since in this case the sub-ensemble
in which the system is definitely located in the restricted region
after the measurement is also pure, and hence its entanglement
can be simply characterised by the entropy of the reduced density
operators. For our spin example we present results showing how the entanglement varies as a function of the parameters of the initial state; for the continuous case, we find also how it depends on the location and size of the chosen regions.  Hence we show that the distribution of entanglement is very different from the distribution of the classical correlations.
\end{abstract}

\pacs{03.67.Mn,03.65.Ud,42.50.Dv}

\maketitle

\section{Introduction}

Studying the entanglement properties of a number of spatially extended
many-body systems including spin chains, coupled fermions, and
harmonic oscillators
\cite{PhysRevA.66.032110,Osterloh.Nature.416.608,vidal:227902,Latorre.QuantumInf.Comput.4.48,Jin.JStatPhys.116.79,Zanardi.JPhysA.35.7947,Martin-Delgado.quantph0207026,PhysRevA.66.042327,Plenio.NewJPhys.6.36,Vedral.CentralEurJPhys.1.289}
has both given information on the potential uses of these systems in
quantum information processing, and yielded insight into their
fundamental properties. Quantum entanglement is a measure of
essentially quantum correlations, and many interacting systems possess
an entangled ground state
\cite{Nielsen.Thesis,Wootters.ContempMaths.305.299,Wootters.JMathPhys.43.4307,PhysRevA.63.052302,PhysRevA.66.032110,Osterloh.Nature.416.608,PhysRevLett.86.910}.

In this paper we address the question: where in configuration space is
the entanglement between two particles located? We pose the question
using the language of spatial entanglement, which plays a significant
role in many physical realizations of QIP (quantum information
processing).  However our results are easily recast in terms of other
types of entanglement.  Specifically, we investigate the location
dependence of the ground-state entanglement between two interacting
subsystems. We choose a pair of coupled harmonic oscillators as an
example, since this is a system for which many exact results are
available \cite{PhysRevA.66.042327,giedke:107901}.  We assign one
oscillator to each of the two communicating parties Alice and Bob, but
perform a thought experiment in which one or both of them first
measure the system in configuration space, with just enough precision
to localise it in some chosen region, and thereafter are restricted to
operations only within that region.  This restriction corresponds to a particular type of projective filtering in configuration space.  We ask how this restriction
affects the spatial entanglement available to them for other purposes---for
example, for teleporting additional qubits between them.  This should be distinguished from the 
approach taken recently by Cavalcanti \textit{et.\ al.} \cite{cavalcanti:062307}, who explored 
the effect of a finite-resolution spatial measurement on the \textit{spin} 
entanglement of a system of noninteracting fermions and also a photonic interferometer. Our research also contrasts with previous studies \cite{botero04,plenio:060503,quantph0607069} of the entanglement of a finite region of space with the rest of the system.

In a previous paper \cite{Concurrence.Density} we investigated the
limiting case where the size of the preliminary-measurement region is
very small, and showed that a smooth two-mode continuous-variable state can be approximated by a pair of qubits and its entanglement fully characterized, even for mixed states, by either \textit{concurrence density} or \textit{negativity density}; here we shall focus on studying the
variations of the entanglement properties with the size of the region.
For the present we assume that the two particles are distinguishable;
the effects of indistinguishability on the phenomena discussed here
are a subject for further work. We argue that the shared entanglement
remaining to Alice and Bob provides a natural measure of where in
configuration space the entanglement was originally located. We show
that the distribution of entanglement is very different from that of
the classical correlations.

\section{Theory}

\subsection{Restricting configuration space by Von Neumann measurements}

Let the configuration space of the whole system be described by the
coordinates $q_{A}$ and $q_{B}$, where $q_{A}$ describes Alice's
particle and $q_{B}$ describes Bob's. We will initially present the
case in which only Alice makes a preparatory measurement on her
system; suppose she has access to some restricted portion $\mathcal{A}$ of the
configuration space of ``her'' particle, whose coordinate is $q_{A}$.
If she measures her system with just enough accuracy to determine
whether it is in region $\mathcal{A}$ or not, but no more, the effect is to
localise the wavefunction either inside, or outside, the chosen
region.  The restriction to lying inside the region corresponds to the
projector
\begin{equation}
\hat{E}_{\mathcal{A}}=\int_{\mathcal{A}}\left|q_{A}\right\rangle \left\langle q_{A}\right|\d q_{A}\otimes\hat{1}_{\mathrm{other}},\label{eq:6.2-1}\end{equation}
 where $\hat{1}_\mathrm{other}$ is the identity operation for all the other
particles (assumed distinguishable) in the system.

\subsubsection{The discarding ensemble}
Suppose $\mathcal{A}$ is of finite extent, and Alice measures the position of her particle with just enough
accuracy to determine whether it is in $\mathcal{A}$ or not. If so, she keeps
the state for further use; if not, she discards it (and tells Bob she
has done so). Then the density matrix appropriate to the ensemble
of retained systems is \begin{eqnarray}
\hat{\rho}_{D,\mathcal{A}} & = & \frac{\hat{E}_{\mathcal{A}}\hat{\rho}\hat{E}_{\mathcal{A}}}{\mathrm{Tr}(\hat{E}_{\mathcal{A}}\hat{\rho})}\label{eq:6.2.1-1}\\
 & = & \frac{\theta_{\mathcal{A}}(q_{A})\rho(q_{A},q_{\mathrm{other}};q'_{A},q'_{\mathrm{other}})\theta_{\mathcal{A}}(q'_{A})}{\int\int_{\mathcal{A}}\rho(q_{A},q_{\mathrm{other}};q_{A},q_{\mathrm{other}})\d q_{A}\d q_{\mathrm{other}}},\nonumber \end{eqnarray}
 where $\theta_{\mathcal{A}}$ is a generalized Heavyside function defined so
that \begin{equation}
\theta_{\mathcal{A}}(q)=\left\{ \begin{array}{l}
1\quad\mathrm{if}\, q\in \mathcal{A}\\
0\quad\mathrm{otherwise}\end{array}\right..\label{eq:6.2.1-2}\end{equation}
 The subscript $D$ refers to the discarding of the unwanted states;
we refer to this density matrix as describing the ``discarding ensemble''.
Note that, if the original $\hat{\rho}$ was a pure state $\left|\psi\right\rangle \left\langle \psi\right|$,
then the post-selected density matrix is also pure: \begin{equation}
\hat{\rho}_{D,\mathcal{A}}=\frac{\hat{E}_{\mathcal{A}}\left|\psi\right\rangle \left\langle \psi\right|\hat{E}_{\mathcal{A}}}{\left\langle \psi\right|\hat{E}_{\mathcal{A}}\left|\psi\right\rangle }.\label{eq:6.2.1-3}\end{equation}
 In particular this means that even though the system has continuous
variables and is therefore infinite-dimensional, its entanglement $\ent_{D,\mathcal{A}}$
is easily calculated through the von Neumann entropy $S(\rhoA)$
of the reduced density matrix $\rhoA=\mathrm{Tr}_B\hat{\rho}$.

\subsubsection{The non-discarding ensemble}
On the other hand if Alice chooses \emph{not} to discard the system when she fails to detect a particle in region $\mathcal{A}$, the
appropriate density matrix is \begin{equation}
\hat{\rho}_{ND}=\hat{E}_{\mathcal{A}}\hat{\rho}\hat{E}_{\mathcal{A}}+\hat{E}_{\mathcal{A}'}\hat{\rho}\hat{E}_{\mathcal{A}'},\label{eq:6.2.1-4}\end{equation}
 where the subscript $ND$ refers to ``non-discarding'' and the complementary projector $\hat{E}_{\mathcal{A}'}$ is defined as \begin{equation}
\hat{E}_{\mathcal{A}'}\equiv\hat{1}-\hat{E}_{\mathcal{A}}=\int_{q_{A}\notin \mathcal{A}}\left|q_{A}\right\rangle \left\langle q_{A}\right|\d q_{A}\otimes\hat{1}_{\mathrm{other}}.\label{eq:6.2.1-5}\end{equation}
 Eq. (\ref{eq:6.2.1-4}) describes a mixed state. It differs from the original density matrix $\hat{\rho}$
in that off-diagonal elements of $\hat{\rho}$ connecting $q_{A}\in \mathcal{A}$
and $q_{A}\notin \mathcal{A}$ have been set to zero.

Let $p_\mathcal{A}=\mathrm{Tr}[\hat{E}_{\mathcal{A}}\hat{\rho}\hat{E}_{\mathcal{A}}]$ be the probability
of finding Alice's particle in $\mathcal{A}$. Since the first and second components
of $\hat{\rho}_{ND}$ can be distinguished by Alice and Bob using
local operations and classical communication (LOCC), they can teleport
$p_\mathcal{A}\ent_{D,\mathcal{A}}+(1-p_\mathcal{A})\ent_{D,\mathcal{A'}}$ qubits on average between them. Hence the distallable entanglement (and therefore
also the entanglement of formation) of $\hat{\rho}_{ND}$ is not less
than $p_\mathcal{A}\ent_{D,\mathcal{A}}+(1-p_\mathcal{A})\ent_{D,\mathcal{A'}}$. On the other hand, Eq. (\ref{eq:6.2.1-4}) also constitutes
a valid decomposition of the non-discarding density matrix $\hat{\rho}_{ND}$
into orthogonal pure states; it follows that the entanglement of formation $\ent_{ND}$
is not greater than the average entanglement of this decomposition:
$\ent_{ND}\le p_\mathcal{A}\ent_{D,\mathcal{A}}+(1-p_\mathcal{A})\ent_{D,\mathcal{A'}}$. The only way these two observations can be consistent
is if \begin{equation}
\ent_{ND}=p_\mathcal{A}\ent_{D,\mathcal{A}}+(1-p_\mathcal{A})\ent_{D,\mathcal{A'}}.\label{eq:6.2.1.-7}\end{equation}

If all the operators available to Alice have support only in region
$\mathcal{A}$ (i.e. if she can neither measure her particle's properties, or
manipulate it in any way, except when it is in $\mathcal{A}$) then the component projected by $\hat{E}_{\mathcal{A}'}$ is ``out of
reach'', and the second
component $\hat{E}_{\mathcal{A}'}\hat{\rho}\hat{E}_{\mathcal{A}'}$ of the state $\hat{\rho}_{ND}$
is functionally equivalent to a separable state as far as any operation that Alice and Bob can perform is concerned. It does not
possess any entanglement properties that are useful to Alice and Bob.  In that case, Eq.\  (\ref{eq:6.2.1.-7}) reduces to $\ent_{ND}=p_\mathcal{A}\ent_{D,\mathcal{A}}$.

\subsubsection{Precise measurements of position\label{sub:Precise-measurements-position}}

If, on the other hand, Alice measures the position accurately, but
again keeps only those occasions when the results lie within $\mathcal{A}$,
the discarding ensemble's density matrix is

\begin{equation}
\hat{\rho}_{P}=\frac{\int_{\mathcal{A}}\hat{E}_{q}\hat{\rho}\hat{E}_{q}\d q}{\int_{\mathcal{A}}\mathrm{Tr}(\hat{E}_{q}\hat{\rho})\d q},\label{eq:6.2.2-1}\end{equation}
 where the subscript $P$ refers to measuring precisely and $\hat{E}_{q}$
is the projector corresponding to measuring Alice's particle $A$
precisely \emph{at} position $q$: \begin{equation}
\hat{E}_{q}=\delta(q_{A}-q).\label{eq:6.2.2-2}\end{equation}
 Eq. (\ref{eq:6.2.2-1}) describes a density matrix that is diagonal
in $q_{A}$; it is a mixed state even if all the measurements where
the particle is not found in $\mathcal{A}$ are discarded. Furthermore, unless
there are some additional degrees of freedom of particle $q_{A}$
which are not measured, the overall density matrix can be written
as an incoherent sum of product states:
\begin{eqnarray}
\hat{\rho}_{P} & = & \int_{\mathcal{A}}\hat{E}_{q}\hat{\rho}\hat{E}_{q}\d q\nonumber \\
 & = & \int_{\mathcal{A}}\left|q_{A}\right\rangle \left\langle q_{A}\right|\d q_{A},\label{eq:6.2.2-3}\end{eqnarray}
 where $\ket{q_A}$ is a state in which particle A is located exactly at $q_A$ and particle B is in some arbitrary state.  $\hat{\rho}_P$ therefore contains no
remaining entanglement with Bob's particle B.
 
Note that in the limit of very small measurement regions, the distinction between precise and imprecise measurements disappears.  The case of vanishingly small regions was analysed in a previous paper \cite{Concurrence.Density}, where it was shown that a well-defined concurrence density exists, and so the concurrence after the measurement is directly proportional to the region size.

\subsubsection{Measurements by both parties}
Exactly analogous formulae can be written down for the cases where
Bob makes a preliminary measurement on his particle, or both partners
make a measurement. In the case where both parties make a preliminary
measurement, the reduced density matrix of Alice's system that is
used to calculate the entanglement will naturally depend also on the
measurement performed by Bob.

\subsubsection{An inequality for the discarding entanglement}
Suppose Alice and Bob divide their configuration spaces into a set of
segments $\mathcal{A}$ and $\mathcal{B}$ respectively, and each make a measurement determining in
which segment the system is located.  In the nondiscarding ensemble,
Eq. (\ref{eq:6.2.1-4}) generalizes to
\begin{equation}
\hat{\rho}_{ND}=\sum_{\mathcal{AB}}\hat{E}_{\mathcal{B}}\hat{E}_{\mathcal{A}}\hat{\rho}\hat{E}_{\mathcal{A}}\hat{E}_{\mathcal{B}},\label{eq:generalizedrhond}
\end{equation}
where 
\begin{equation}
\sum_{\mathcal{A}}\hat{E}_{\mathcal{A}}=\sum_{\mathcal{B}}\hat{E}_{\mathcal{B}}=\hat{1}.
\end{equation}
However, this corresponds to a local operation performed by Alice and
Bob.  Their shared entanglement is non-increasing under this
operation; therefore, 
\begin{equation}
\ent(\hat{\rho})\ge \ent(\hat{\rho}_{ND}).\label{eq:inequality}
\end{equation}
But, by a straightforward extension of the argument given above, 
\begin{equation}
\ent(\hat{\rho}_{ND})=\sum_{\mathcal{AB}}p_{\mathcal{AB}}\ent_D(\hat{\rho}_{D,\mathcal{A}\mathcal{B}}),\label{eq:erelate}
\end{equation}
where
\begin{equation}
p_{\mathcal{AB}}=\mathrm{Tr}[\hat{E}_{\mathcal{B}}\hat{E}_{\mathcal{A}}\hat{\rho}]
\end{equation}
is the probability of finding Alice's part of the system in
$\mathcal{A}$ and Bob's part in $\mathcal{B}$, and 
\begin{equation}
\hat{\rho}_{D,\mathcal{A}\mathcal{B}}=\frac{\hat{E}_{\mathcal{B}}\hat{E}_{\mathcal{A}}\hat{\rho}\hat{E}_{\mathcal{A}}\hat{E}_{\mathcal{B}}}{p_{\mathcal{AB}}}
\end{equation}
is the density matrix in the discarding ensemble after this
measurement result has been obtained.  Combining Eq.
(\ref{eq:inequality}) and Eq. (\ref{eq:erelate}) we obtain the following
inequality for the average of the entanglement in the discarding
ensemble over all the partitions:
\begin{equation}
\sum_{\mathcal{AB}}p_{\mathcal{AB}}\ent_D(\hat{\rho}_{D,\mathcal{A}\mathcal{B}})
\le \ent(\hat{\rho}).\label{eq:discardinginequality}
\end{equation}

\subsection{Spin systems}

We can make an exactly analogous theory for the case where Alice and
Bob share a system defined on some other state space, for example
a spin system---perhaps more familiar in quantum information theory.
We simply replace the projection operator $\hat{E}_{\mathcal{A}}$ by one defined
in spin space; for example, $\hat{E}_{\mathcal{A}}$ might project onto states with
a specified spin component in a given direction. The rest of the theory
is as outlined above.

\section{The physical systems}

\subsection{Spins}

Suppose that both Alice and Bob each possess two spins; the first
spins belonging to each of them are entangled, as are the second spins,
and the overall state $\left|\psi\right\rangle $ of the system is
a product of the state of the two pairs. For example, we could write

\begin{eqnarray}
\left|\psi\right\rangle  & = & (\cos\theta_{1}\left|\uparrow_{A1}\uparrow_{B1}\right\rangle +\sin\theta_{1}\left|\downarrow_{A1}\downarrow_{B1}\right\rangle )\nonumber \\
 &  & \quad\otimes(\cos\theta_{2}\left|\uparrow_{A2}\uparrow_{B2}\right\rangle +\sin\theta_{2}\left|\downarrow_{A2}\downarrow_{B2}\right\rangle );\label{eq:spins-1}\end{eqnarray}
 the state is pure so entanglement between Alice's and Bob's subsystems
is well quantified by the von Neumann entropy of the reduced density
matrix. Suppose also that Alice and Bob can only handle systems if
the total spins $M_{s}$ available to each party are such that $M_{s}=0$;
perhaps the parts of the state with non-zero moment are lost because
of the presence of large fluctuating fields in the environment. In
the discarding ensemble defined by this restriction, the state becomes
\begin{eqnarray}
\left|\psi\right\rangle _{R} & = & \frac{1}{\sqrt{1-\cos2\theta_{1}\cos2\theta_{2}}}\nonumber \\
 &  & \quad(\cos\theta_{1}\left|\uparrow_{A1}\uparrow_{B1}\right\rangle \sin\theta_{2}\left|\downarrow_{A2}\downarrow_{B2}\right\rangle \nonumber \\
 &  & \quad+\sin\theta_{1}\left|\downarrow_{A1}\downarrow_{B1}\right\rangle \cos\theta_{2}\left|\uparrow_{A2}\uparrow_{B2}\right\rangle ),\label{eq:spins-2}\end{eqnarray}
 again this restricted state is pure but entanglement should become
quite different.

This type of measurement is familar in other
contexts---for example entanglement distillation and concentration \cite{PhysRevLett.76.722,PhysRevA.53.2046}.

Entanglement of a bi-partite mixed spin state can also be easily quantified
by using negativity $\mathcal{N}(\hat{\rho})$ instead as the entanglement
measure. We define negativity as the sum of the magnitudes of the
negative eigenvalues $\lambda_{i}$ of the partially transposed density
matrix $\hat{\rho}^{T_{B}}$, \begin{equation}
\mathcal{N}(\hat{\rho})=\sum_{i\,\textbf{s.t.}\,\lambda_{i}<0}|\lambda_{i}|.\label{eq:DefNegativity}\end{equation}

Consider the mixed state defined by
\begin{equation}\label{eq:mixedstatedef}
\hat{\rho}=\frac{16F-1}{15}\left|\psi\right\rangle \left\langle \psi\right|+\frac{1-F}{15}\hat{1},
\end{equation}
where $\left|\psi\right\rangle $ is as defined in Eq. (\ref{eq:spins-1}) 
(in contrast to the definition of Werner states, this is not a maximally
entangled state) and $F\in[1/16,1]$. Note that when $F=1$, the state becomes pure. Again an example of discarding ensembles can be obtained by projecting the state (\ref{eq:mixedstatedef}) onto $M_{s}=0$ subspace and renormalizing accordingly.

\subsection{Harmonic oscillators\label{sub:Harmonic-oscillators}}

The density matrix of a Gaussian state can be written in the coordinate
representation \cite{PhysRevA.36.3868} as

\begin{eqnarray}
\left\langle q\left|\hat{\rho}\right|q'\right\rangle  & \equiv & \rho(q;q')\nonumber \\
 & = & \zeta_{1}\mathrm{exp}[-q^{T}\mathbf{L}q-q'^{T}\mathbf{L}q'\nonumber \\
 &  & \qquad-\frac{1}{2}(q-q')^{T}\mathbf{M}(q-q')\nonumber \\
 &  & \qquad+\frac{i}{2}(q-q')^{T}\mathbf{K}(q+q')],\label{eq:2-6}\end{eqnarray}
 where $\zeta_{1}$ is a normalization constant, and where $\mathbf{L},$
$\mathbf{M}$ and $\mathbf{K}$ are real $N$-dimensional matrices
with $\mathbf{L}$ and $\mathbf{M}$ symmetric, while $\mathbf{K}$
is arbitrary. These matrices are related to the covariance matrix
$\gamma$ by \begin{equation}
\frac{1}{2}\gamma^{-1}=\left(\begin{array}{cc}
\boldsymbol{1} & 0\\
-\mathbf{K} & \boldsymbol{1}\end{array}\right)^{T}\left(\begin{array}{cc}
2\mathbf{L} & 0\\
0 & \frac{1}{2}(\mathbf{L}+\mathbf{M})^{-1}\end{array}\right)\left(\begin{array}{cc}
\boldsymbol{1} & 0\\
-\mathbf{K} & \boldsymbol{1}\end{array}\right).\label{eq:2-8}\end{equation}
We note that for a pure state, $\mathbf{M}=0$ and \textbf{$\mathbf{K}$}
is symmetric.

Consider a harmonic system with a Hamiltonian  (taking $\hbar=1$)  \begin{equation}
\hat{H}=R^{T}(\begin{array}{cc}
\mathbf{V}m\omega^{2}/2 & 0\\
0 & \boldsymbol{1}_{N}/(2m)\end{array})R,\label{eq:2-1}\end{equation}
 where the vector $R$ of quadrature operators is given by the positions
$R_{j}=\hat{X}_{j}$ and conjugate momenta $R_{N+j}=\hat{P}_{j},$
for $1\leq j\leq N,$, the $N\times N$ matrix $\mathbf{V}$ contains
the coupling coefficients, and $\omega$ is the natural frequency
of uncoupled oscillators. For a translationally invariant system the
potential matrix elements depend only on the difference between the
indices: $\mathbf{V}_{j,k}=v_{(j-k)\mathrm{mod\,}N}$ for $1\leq j,k\leq N$.
The covariance matrix of the ground state is then \cite{PhysRevA.66.042327}
\begin{eqnarray}
\gamma & = & \frac{1}{2}(\frac{\gamma_{x}}{m\omega}\oplus m\omega\gamma_{p})\nonumber \\
 & = & \frac{1}{2}(\frac{\mathbf{V}^{-1/2}}{m\omega}\oplus m\omega\mathbf{V}^{1/2}).\label{eq:2-5}\end{eqnarray}
Since the Hamiltonian given in Eq. (\ref{eq:2-1}) has
no coupling between position and momentum variables, $\gamma$
is block diagonal and hence $\mathbf{K}=0$. Furthermore if there
are only nearest-neighbour interactions, with a Hooke's-law spring constant $K$, the interaction
strength is characterized by the single dimensionless parameter \begin{equation}
\alpha=\frac{2K}{m\omega^{2}}.\label{eq:2-4}\end{equation}

For the two-oscillator ground state we therefore have only one non-zero matrix:
\begin{equation}
\mathbf{L}=\frac{m\omega}{8}\left(\begin{array}{cc}
1+\sqrt{1+4\alpha} & 1-\sqrt{1+4\alpha}\\
1-\sqrt{1+4\alpha} & 1+\sqrt{1+4\alpha}\end{array}\right).\label{eq:2-9}\end{equation}
The 1-particle reduced density matrices can then be easily obtained
by quadrature; for Particle 1,
\begin{eqnarray}
\rhoA(q_{A};q_{A}^{\prime}) & = & \int_{-\infty}^{\infty}\d q_{B}\rho(q_{A},q_{B};q_{A}^{\prime},q_{B})\label{eq:2-11}\\
 & = & \sqrt{\frac{2C_{1}-2C_{2}}{\pi}}\nonumber \\
 &  & \quad\mathrm{exp}[-C_{1}(q_{A}^{2}+q_{A}^{\prime2})+2C_{2}q_{A}q_{A}^{\prime}],\nonumber \end{eqnarray}where the state is normalized to unity and the constants $C_1$ and $C_2$ are 
\begin{equation}C_{1}=\frac{1+2\alpha+3\sqrt{1+4\alpha}}{8+8\sqrt{1+4\alpha}}m\omega\label{eq:2-12}\end{equation}
 and
 \begin{equation}
C_{2}=\frac{\alpha(\sqrt{1+4\alpha}-1)}{8(1+2\alpha+\sqrt{1+4\alpha})}m\omega.\label{eq:2-13}\end{equation}
From Eq. (\ref{eq:2-11}), we can also define the Gaussian characteristic
length $\sigma$ which characterizes the probability distribution of
a single particle:
\begin{equation}
\sigma=\left[2m\omega\left(\frac{\sqrt{1+4\alpha}}{1+1\sqrt{1+4\alpha}}\right)\right]^{-\frac{1}{2}}.\label{eq:2-14}
\end{equation}

For bipartite Gaussian states, the entanglement of
formation is known exactly \cite{rendell:012330}.  For the ground state of our
system, the value is
\begin{equation}
S(\rhoA)=-\log_2{(1-w)}-\frac{w\log_2 w}{(1-w)},\label{eq:gaussianeof}
\end{equation}
where 
\begin{equation}
w=\frac{1+3\,{\sqrt{1+4\,\alpha}}+2\,[\alpha-\,{\left(1+4\,\alpha\right)}^{\frac{1}{4}}-\,{\left(1+4\,\alpha\right)}^{\frac{3}{4}}]}{1+2\,\alpha-{\sqrt{1+4\,\alpha}}}.\end{equation}

\section{Method\label{sec:Method}}
For the spin system, the entanglement can be calculated straightforwardly by standard methods; for the two-oscillator system, 
we calculate the von Neumann entropy $S(\rhoA)$, and hence the entanglement, numerically
by using two different approaches.

\subsection{Expansion in a complete set}

We define an orthonormal set of functions, $\{\phi_{n}(q)\}$, with
support in a region $\mathcal{A}$ of configuration space of width $2a$ centred
at coordinate $\bar{q}$: \begin{equation}
\int_{\bar{q}-a}^{\bar{q}+a}\phi_{n}(q)\phi_{m}^{*}(q)=\delta_{nm}.\label{eq:TA-2}\end{equation}
 A suitable choice is \begin{eqnarray}
\phi_{n}(q) & = & \sqrt{\frac{1}{a}}\cos(\frac{(q-\bar{q})n\pi}{2a})\quad\textrm{$n$ is odd}\nonumber \\
\phi_{n}(q) & = & \sqrt{\frac{1}{a}}\sin(\frac{(q-\bar{q})n\pi}{2a})\quad\textrm{$n$ is even},\label{eq:TA-1}\\
 & = & 0\quad\textrm{if}\quad\left|q-\bar{q}\right|>a\nonumber \end{eqnarray}

We then approximate the appropriate post-selected density matrix by
an expansion in a finite set of the functions defined in Eq. (\ref{eq:TA-1});
as an example, if only Alice makes a preliminary measurement to localise her particle in the region $\mathcal{A}$, we
have: \begin{equation}
\rhoA_{D}(q_{A};q_{A}^{\prime})=\sum_{mn}^{N}\rho_{mn}\phi_{m}(q_{A})\phi_{n}^{*}(q_{A}^{\prime}),\label{eq:TA-3}\end{equation}
 with $\rho_{mn}$ given by \begin{eqnarray}
\rho_{mn} & = & \int_{\bar{q}_{A}-a}^{\bar{q}_{A}+a}\d q_{A}\int_{\bar{q}_{A}-a}^{\bar{q}_{A}+a}\d q_{A}^{\prime}\nonumber \\
 &  & \quad\phi_{m}^{*}(q_{A})\rhoA(q_{A};q_{A}^{\prime})\phi_{n}(q_{A}^{\prime}),\label{eq:TA-4}\end{eqnarray}
 where $\rhoA(q_{A};q_{A}^{\prime})$ is Eq. (\ref{eq:2-11}).
We normalize $\rhoA_{D}(q_{A};q_{A}^{\prime})$ by its trace and can
then quantify entanglement by calculating the von Neumann entropy
from this normalized $\rhoA_{D}(q_{A};q_{A}^{\prime})$. Unfortunately
the quadratures in Eq. (\ref{eq:TA-4}) must be performed numerically,
making this approach relatively time-consuming.

\subsection{Configuration-space grid}

We therefore explored also a direct real-space approach, in which
we first \emph{discretize} the configuration space into a finite number
of measurement ``bins'', then select only those bins that correspond
to the regions within which Alice's and Bob's respective particles localise.
For example, consider again the case in which only Alice makes a preliminary measurement, if the region is $\bar{q}_{A}-a\le q_{A}\le\bar{q}_{A}+a$,
we divide this space into $N_B$ regions with $N_B+1$ equally spaced
points ($q_{A}$'s) covering the intervals from $q_{A}=\bar{q}_{A}-a$
to $\bar{q}_{A}+a$. We then build the $(N_B+1)\times(N_B+1)$ post-selected 1-particle reduced density matrix $\rhoA_{D}(q_{A};q_{A}^{\prime})$
by calculating its elements $\rho_{mn}$'s from the 1-particle reduced
density matrix Eq. (\ref{eq:2-11}): \begin{equation}
\rho_{mn}=\rhoA(q_{A}^{m};q_{A}^{n})\quad\mathrm{for}\quad1\leq m,n\leq N_B+1.\label{eq:DDM-1}\end{equation}
 As in the other approach, we calculate the von Neumann entropy of
the normalized $\rhoA_{D}(q_{A};q_{A}^{\prime})$ in order to quantify
the entanglement.

Note that if on the other hand both parties make a preliminary measurement,
we start from the full 2-particle density matrix and apply Bob's restrictions
w.r.t his oscillator before we reduce it into the 1-particle density
matrix for Alice's oscillator. 

We find that results from the two approaches converge to the same
values as the number of grid points, or the number of expansion functions,
tend to infinity. Since the second (grid-based) approach is much more
efficient to compute, it has been used for all the results
presented in this paper.

\section{Results}

\subsection{The spin system}

\begin{figure*}
\subfigure[Unrestricted]{\includegraphics[width=0.33\linewidth]{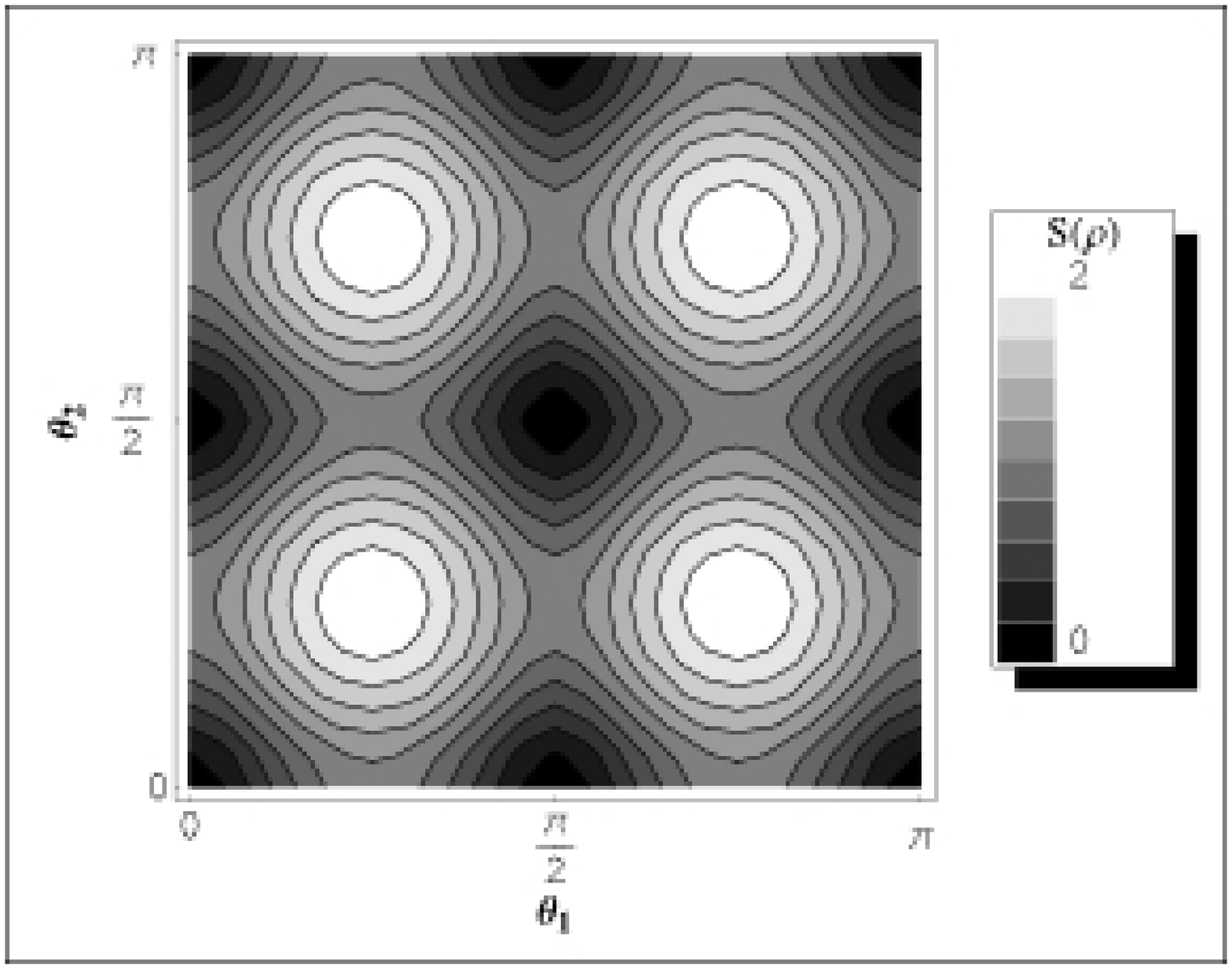}}\subfigure[Restricted (discarding ensemble)]{\includegraphics[width=0.33\linewidth]{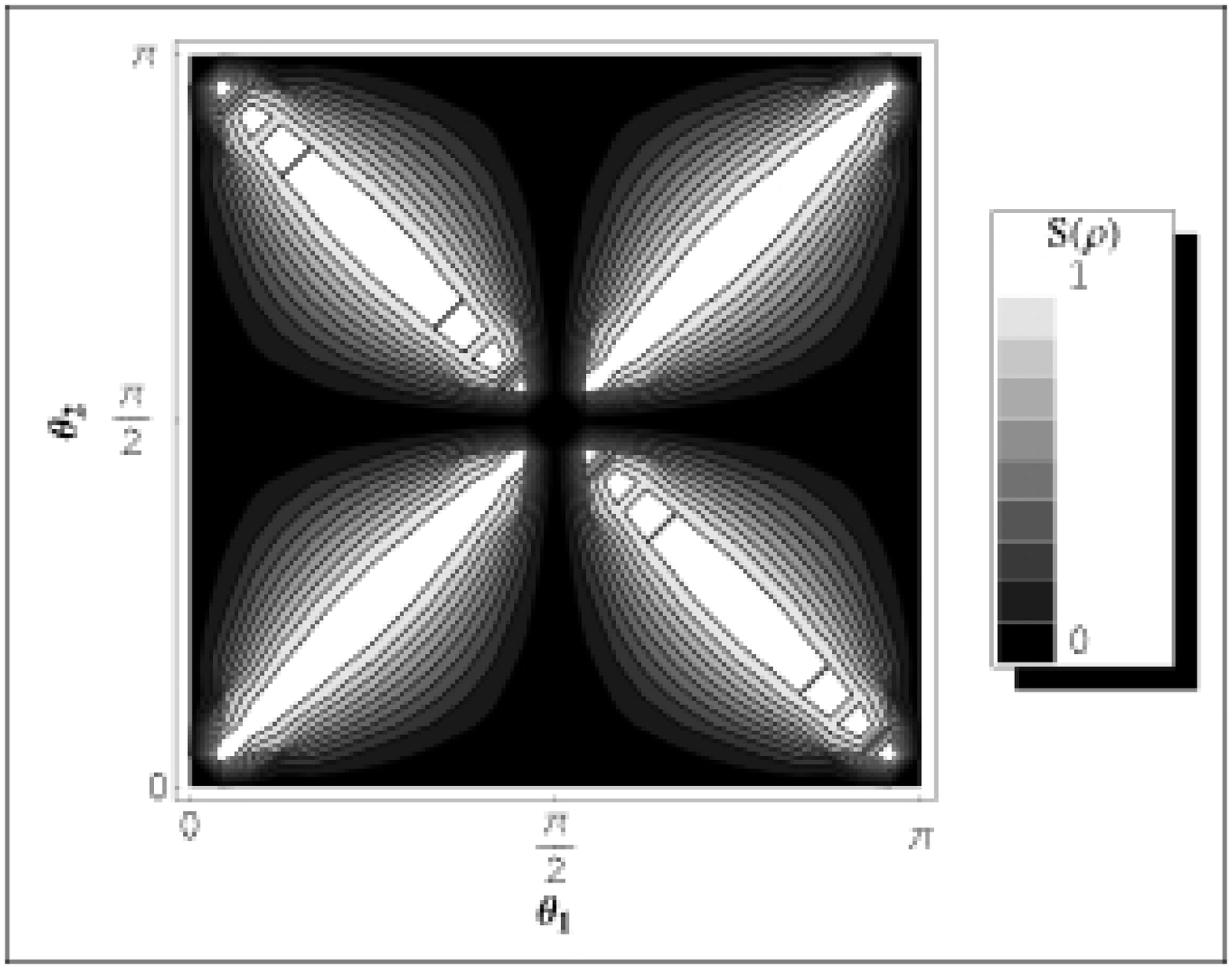}}\subfigure[Differences]{\includegraphics[width=0.33\linewidth]{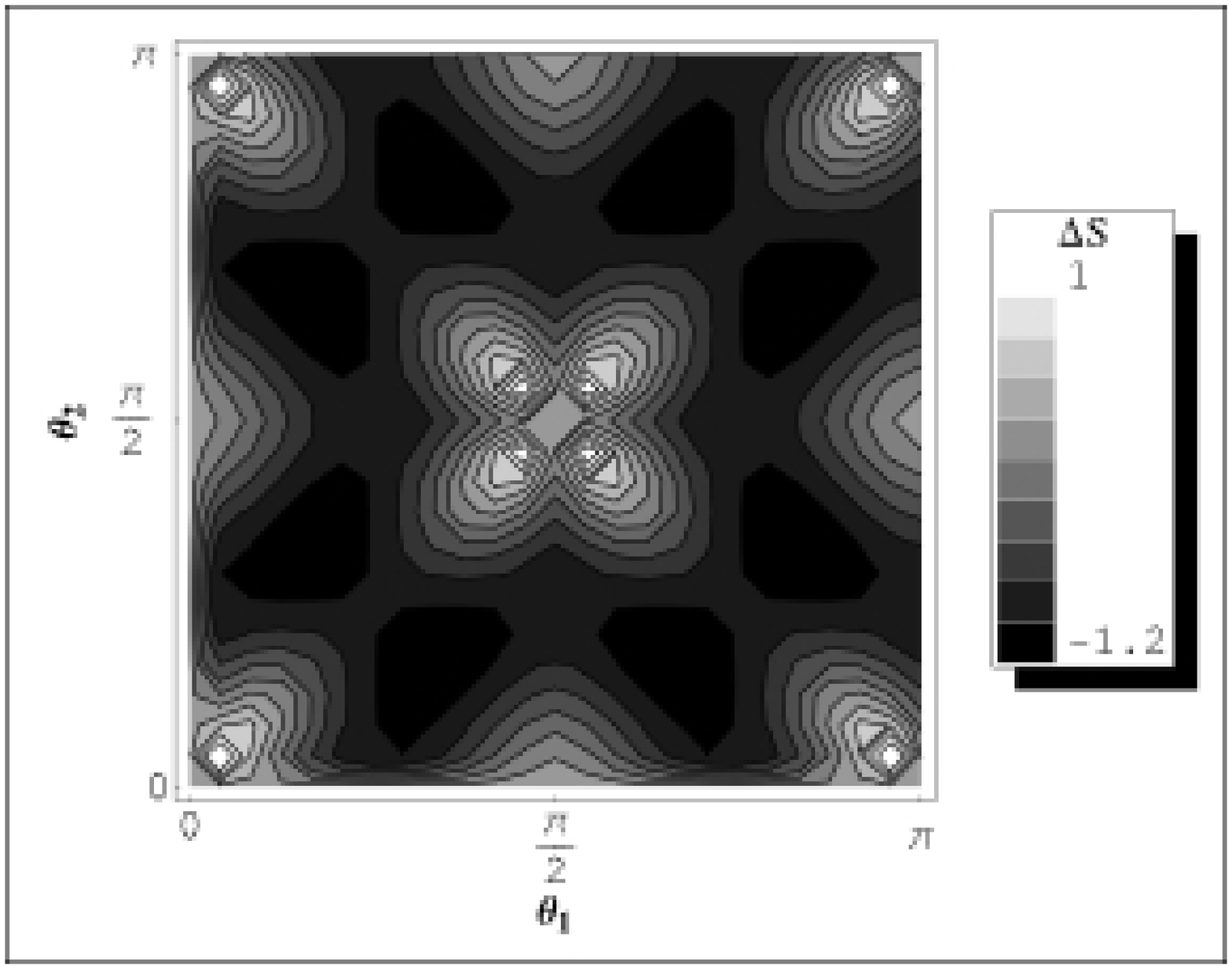}}

\caption{Entanglement (von Neumann entropy $S(\rhoA)$) present in the chosen
spin system (a) when the total spins $M_{s}$ is unrestricted,
(b) in the discarding ensemble when $M_{s}$ for each party must be 0. (c)
Entanglement differences between the two cases; $\Delta S=S(\rhoA_D)-S(\rhoA)$.
\label{fig:Spins}}
\end{figure*}

\begin{figure*}
\begin{centering}(A) $F=0.3$\par\end{centering}

\includegraphics[width=0.33\linewidth,keepaspectratio]{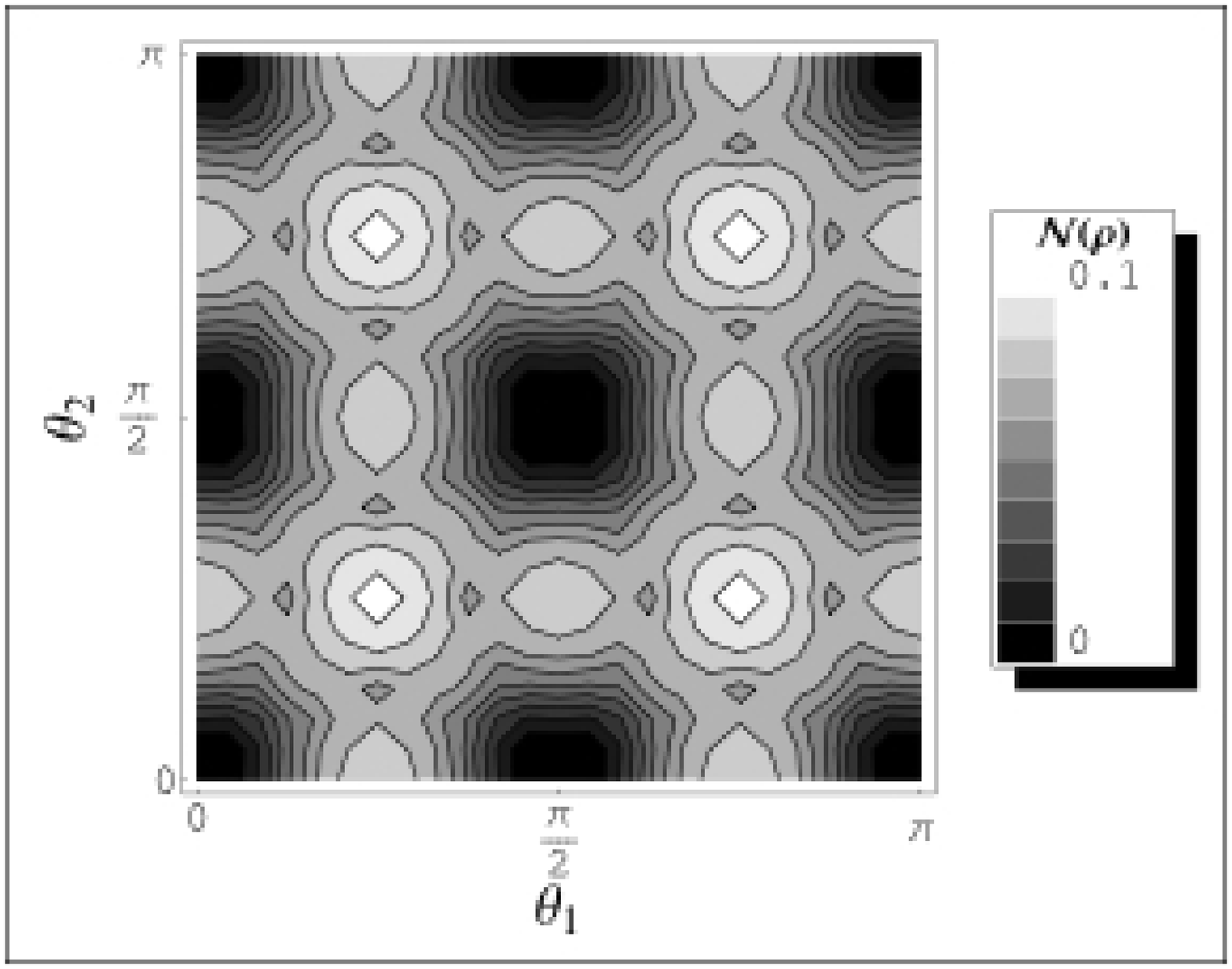}\includegraphics[width=0.33\linewidth,keepaspectratio]{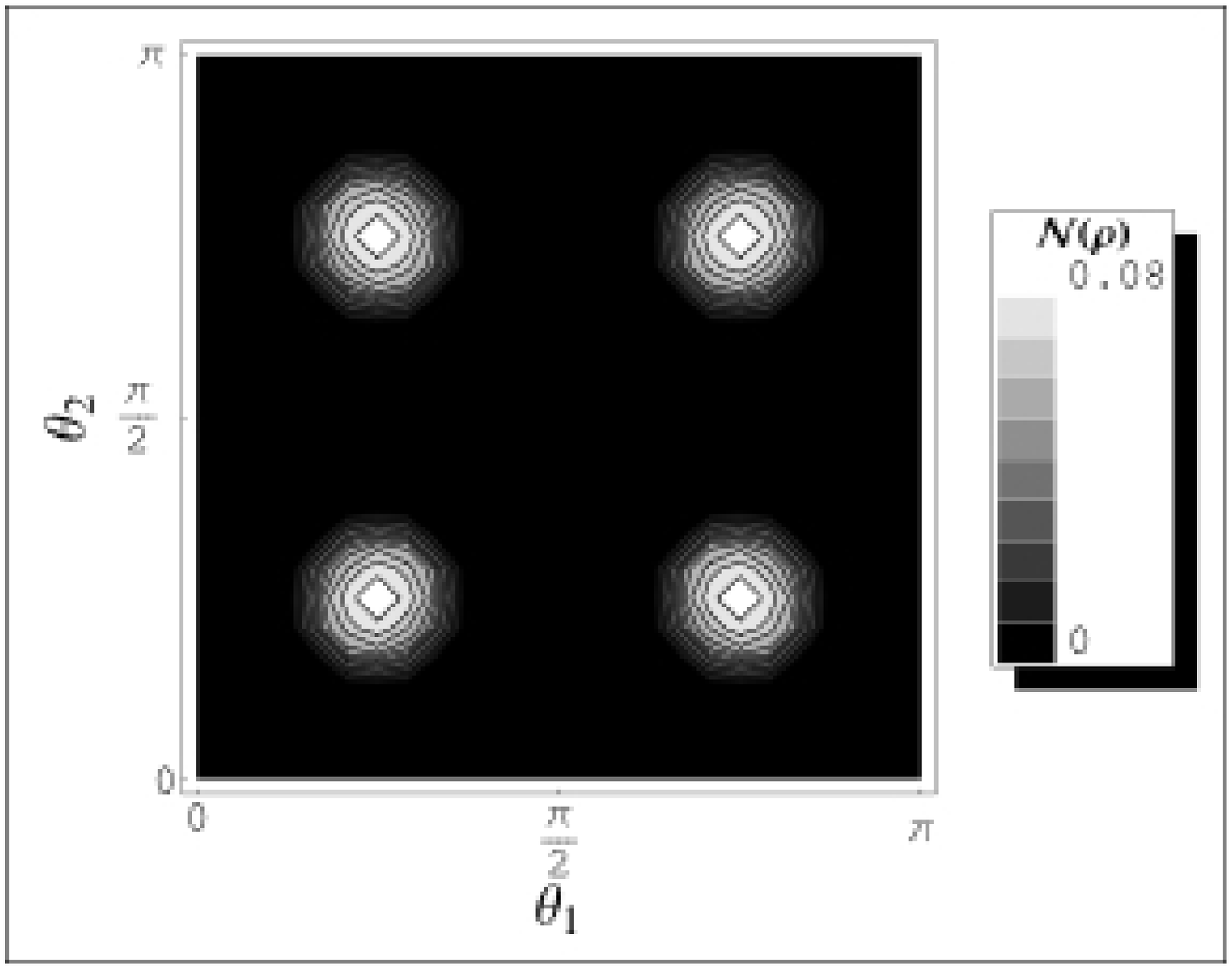}\includegraphics[width=0.33\linewidth,keepaspectratio]{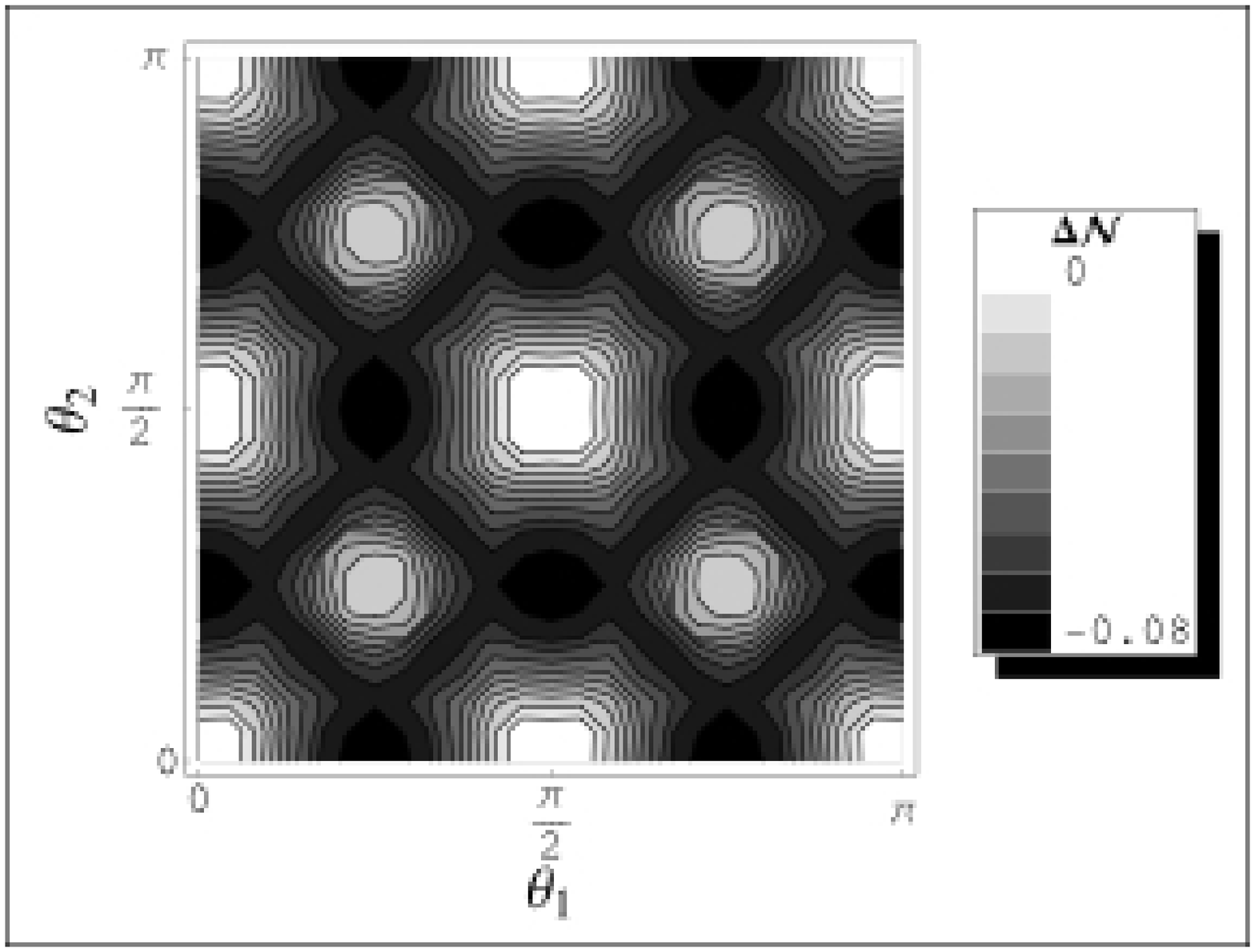}

\begin{centering}(B) $F=0.65$\par\end{centering}

\includegraphics[width=0.33\linewidth,keepaspectratio]{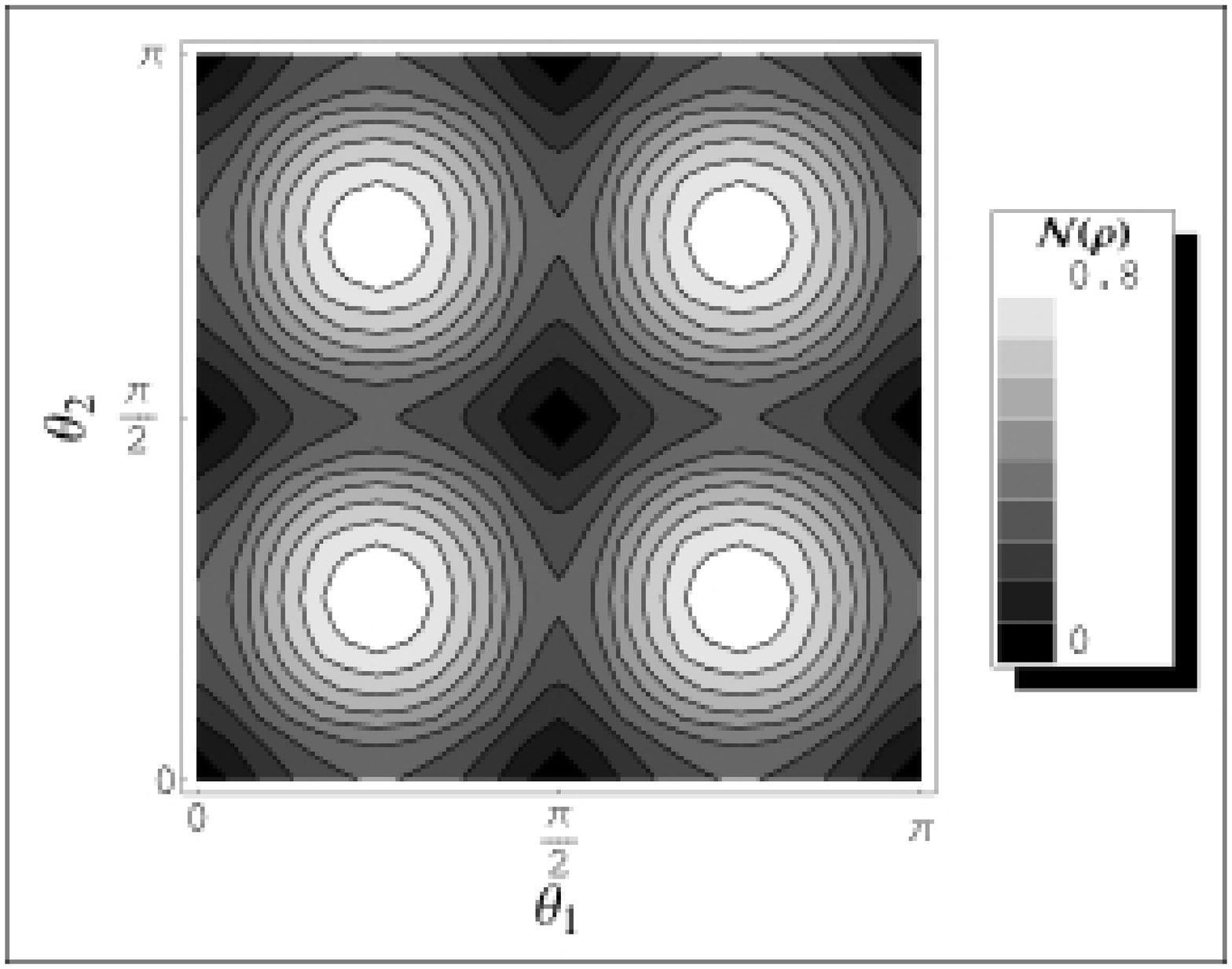}\includegraphics[width=0.33\linewidth,keepaspectratio]{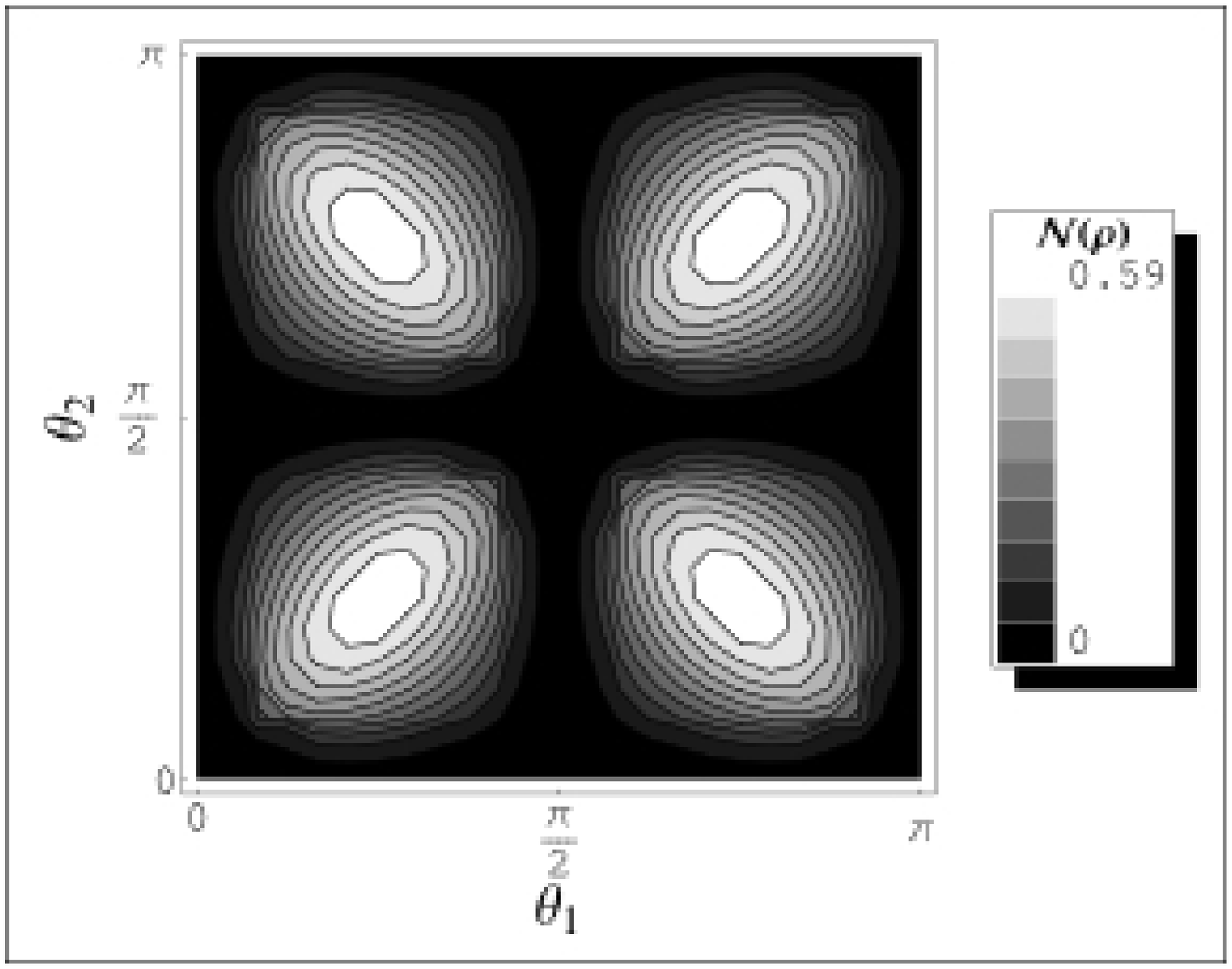}\includegraphics[width=0.33\linewidth,keepaspectratio]{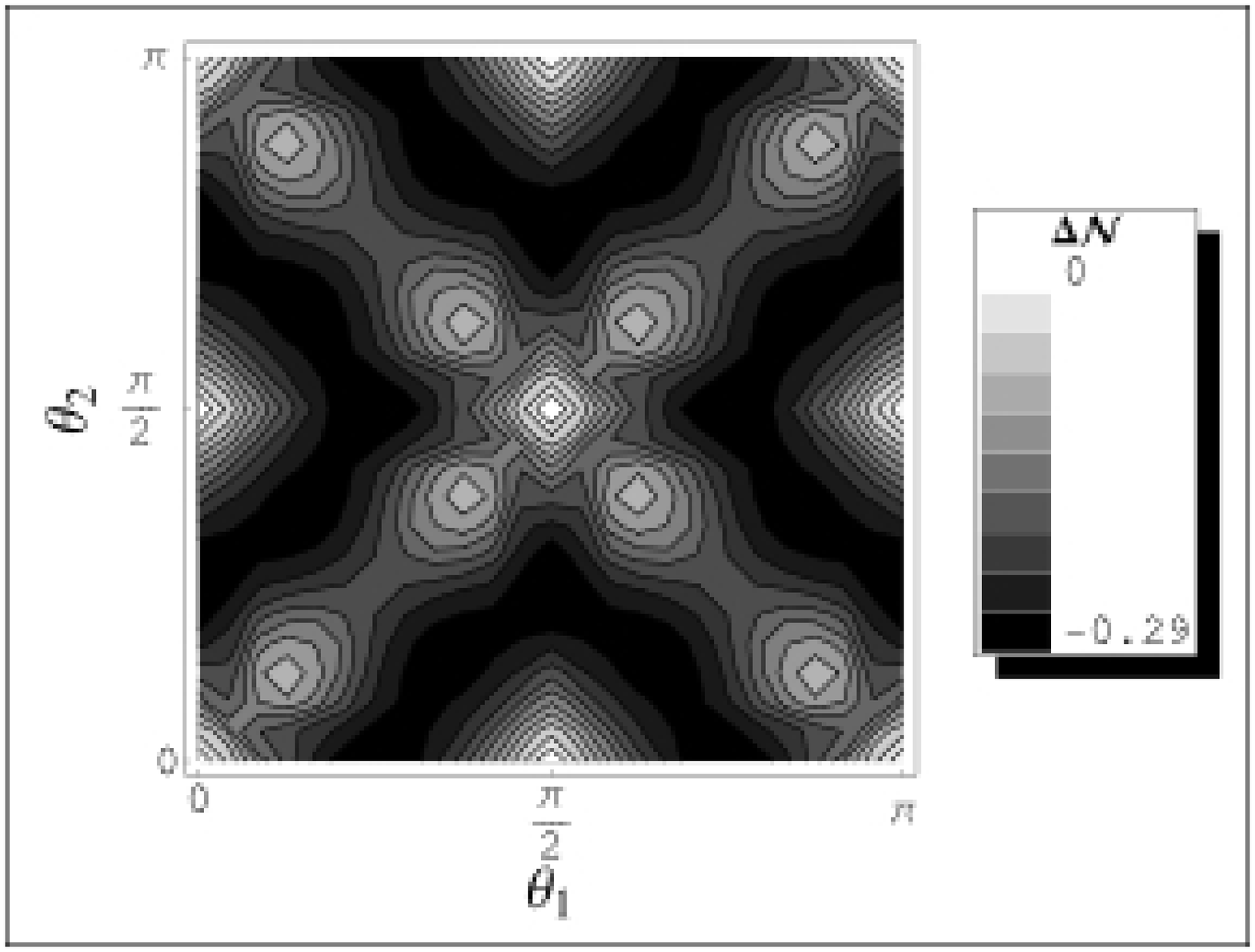}

\begin{centering}(C) $F=1.0$\par\end{centering}

\subfigure[Unrestricted]{\includegraphics[width=0.33\linewidth,keepaspectratio]{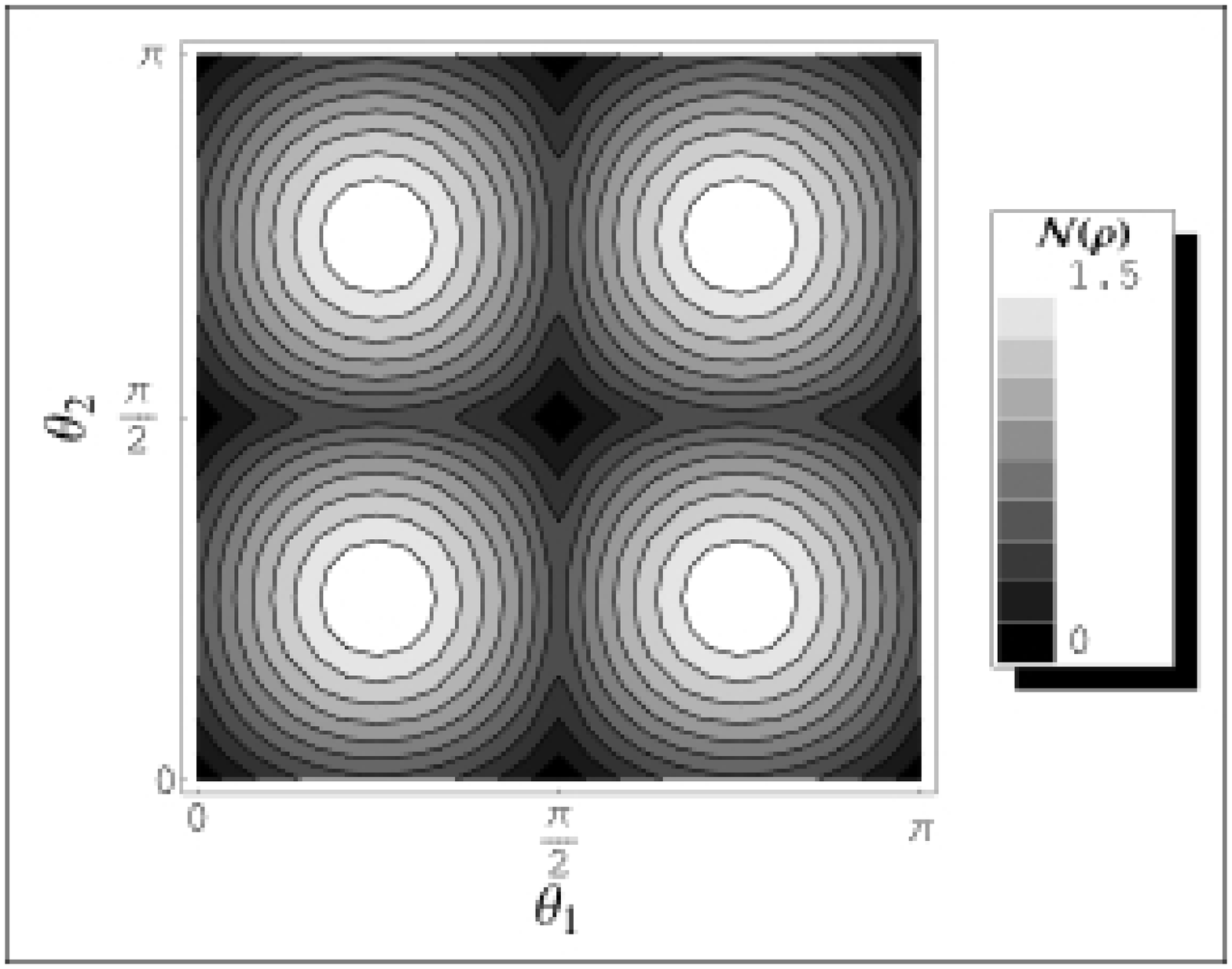}}\subfigure[Restricted (discarding ensemble)]{\includegraphics[width=0.33\linewidth,keepaspectratio]{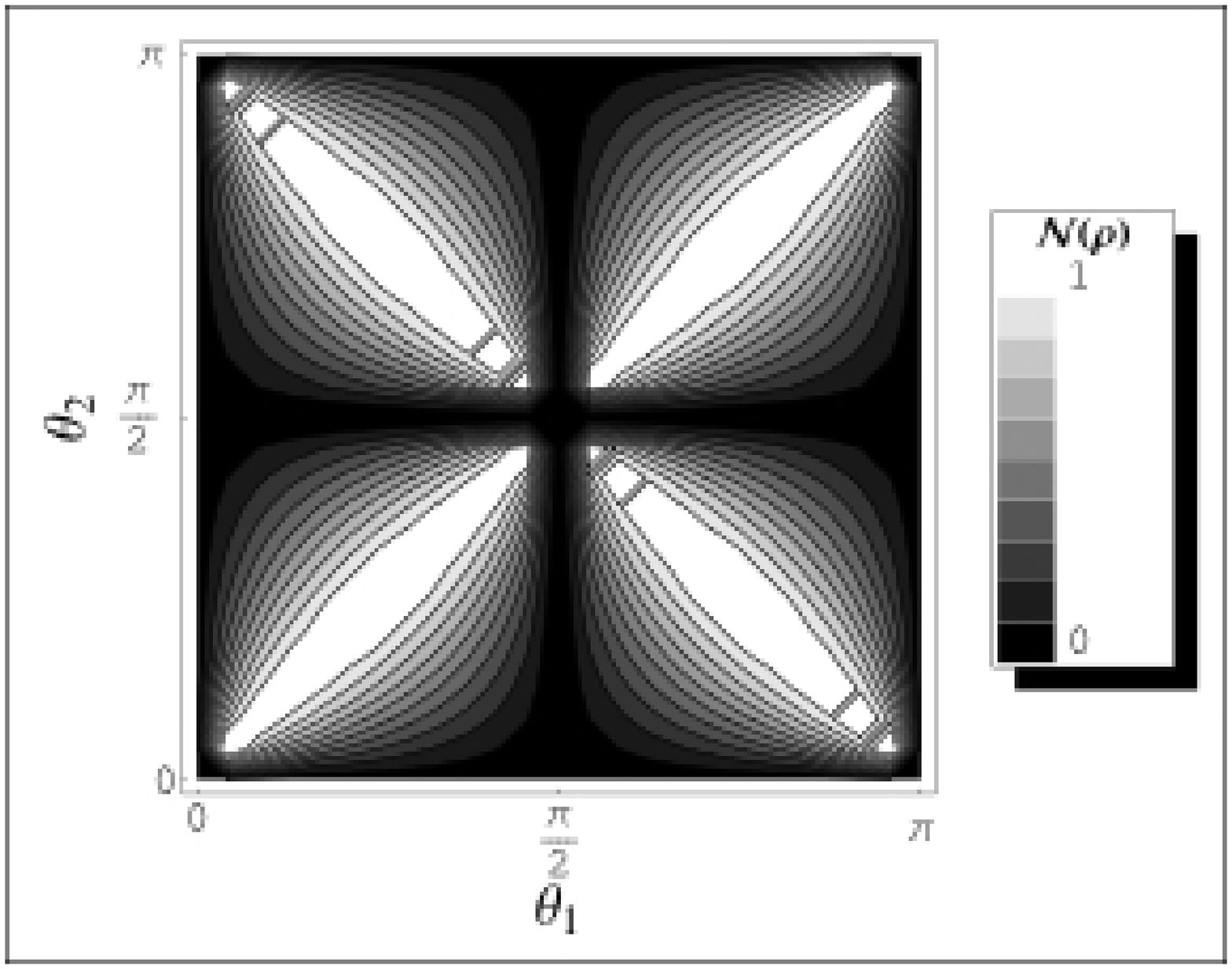}}\subfigure[Differences]{\includegraphics[width=0.33\linewidth,keepaspectratio]{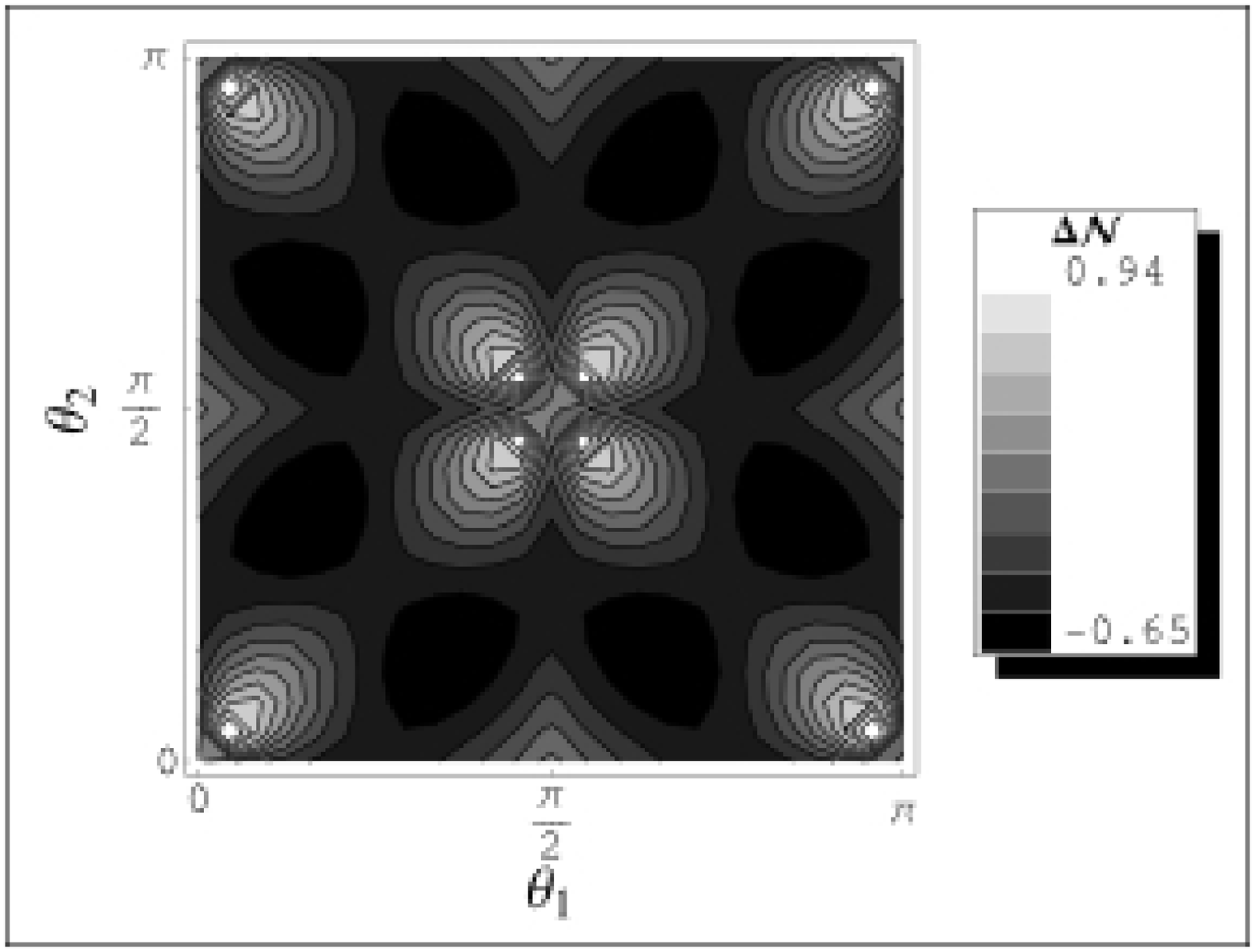}}

\caption{Entanglement (negativity $\mathcal{N}(\hat{\rho})$) present in the mixed state (\ref{eq:mixedstatedef}) (a) when the total spins $M_{s}$ is unrestricted, (b)
in the discarding ensemble when $M_{s}$ for each party must be 0.
(c) Entanglement differences between the two cases; $\Delta\mathcal{N}=\mathcal{N}(\hat{\rho}_{D})-\mathcal{N}(\hat{\rho}_{o})$.
$F$ determines the ``mixedness'' of the state; when $F=1$, the
state is pure. \label{fig:MixedSpins}}
\end{figure*}

\begin{figure}
\includegraphics[width=0.8\linewidth,keepaspectratio]{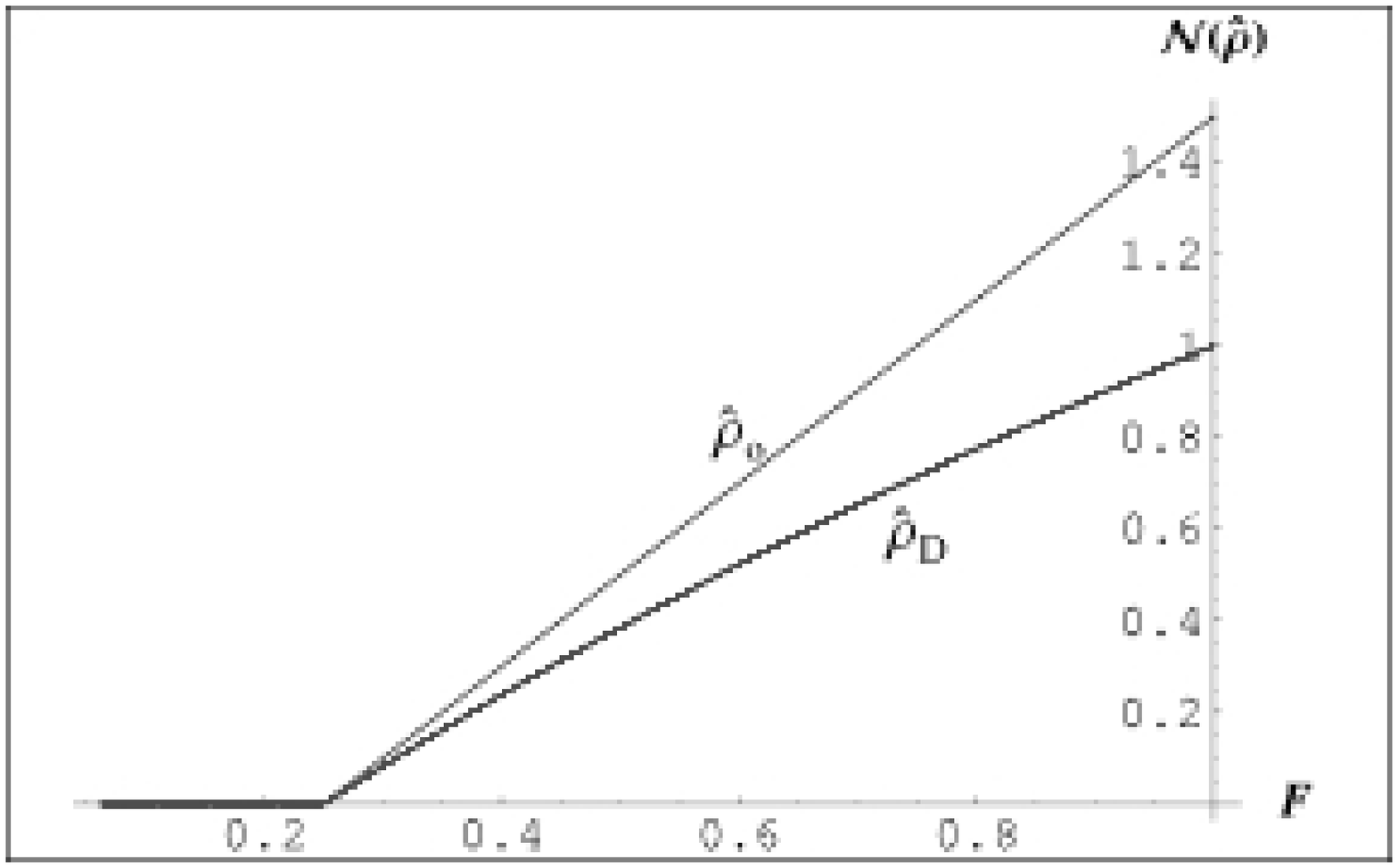}

\caption{Variation of the entanglement (negativity $\mathcal{N}(\hat{\rho})$)
with $F\in[1/16,1]$. $F$ is a quantity that determines the mixedness
of the state as defined by Eq. (\ref{eq:mixedstatedef}). The solid line is for the original ensemble $\hat{\rho}_{o}$ whereas
the dashed line is for the corresponding discarding ensemble $\hat{\rho}_{D}$. Both
$\theta_{1}$ and $\theta_{2}$ have been set to $\pi/4$ to produce
the plots.\label{fig:NegativityWithF}}
\end{figure}

\subsubsection{Pure states}

We present results in Fig. \ref{fig:Spins}. For the spin system we
consider, entanglement present in the state Eq. (\ref{eq:spins-1})
depends on $\theta_{1}$ and $\theta_{2}$ with periods of $\pi/2$,
as shown in Fig. \ref{fig:Spins}(\textbf{a}). The maximum entanglement
is 2 ebits and occurs when both pairs of spins are in the Bell state
($\theta_{1}=\theta_{2}=(2n+1)\pi/4$). When $\theta_{1}=\theta_{2}=n\pi/2$,
the state reduces to all spins either all up or down so completely
loses any entanglement.

Now if the restricted region for both Alice and Bob is chosen to be the
subspace in which the total $z$-component of spin takes the value
zero, and we work in the discarding ensemble so all other states are
eliminated, the entanglement properties of the system become very
different. Fig. \ref{fig:Spins}(\textbf{b}) shows that the entanglement
distribution of the restricted state has periods of $\pi$ instead
of $\pi/2$, and the maximum possible entanglement (now 1 ebit since
the restricted subspaces for both Alice and Bob are two-dimensional)
is achieved whenever $\theta_{1}=\theta_{2}$ or $\theta_{1}+\theta_{2}$
are integer multiples of $\pi$ so that the restricted state is in
the Bell state. Note that there is a singularity whenever $\cos2\theta_{1}\cos2\theta_{2}=1$.

If we compare Fig. \ref{fig:Spins}(\textbf{a}) and Fig. \ref{fig:Spins}(\textbf{b}),
it seems that in some instances the restricted state has higher entanglement.
This is indeed the case as shown in Fig. \ref{fig:Spins}(\textbf{c}),
where $\Delta S=S(\rhoA_D)-S(\rhoA)$
is plotted against both $\theta_{1}$ and $\theta_{2}$. This is an
example of the familiar process of entanglement concentration \cite{PhysRevLett.76.722,PhysRevA.53.2046},
in which some partial entanglement is concentrated after chosen local
measurements. Entanglement is not created on average in
our example because the probability of finding $M_{s}=0$ is not 100\%.  Therefore the inequality Eq. (\ref{eq:discardinginequality}) is not violated.

\subsubsection{Mixed states}

Now we perform a similar calculation for the mixed state (\ref{eq:mixedstatedef}), comparing the entanglement (as quantified by the negativity) present when the total spins
$M_{s}$ is unrestricted and the entanglement in the discarding ensemble
when $M_{s}$ for each party must be 0. The results are presented
in Fig. \ref{fig:MixedSpins}. We choose three values of $F$ for
comparison; $F=0.3,$ $F=0.65$ and $F=1$. 

In Fig. \ref{fig:NegativityWithF}, we plot the variation of the
entanglement $\mathcal{N}(\hat{\rho})$ with $F$ by choosing both $\theta_{1}$
and $\theta_{2}$ to be $\pi/4$ (other values can be chosen without
affecting the qualitative behaviour but entanglement will not vanish at
smaller values of $F$). The case when the total $M_{s}$ is
unrestricted $\hat{\rho}_{o}$ is plotted as a solid line, while the case of the corresponding
discarding ensemble $\hat{\rho}_{D}$ is plotted as a dashed line.  The entanglement in both cases 
vanishes at $F=0.25$. This is similar to what we observed in a previous paper \cite{Concurrence.Density}: in that case, we showed that the entanglement (as quantified by the negativity) of a two-mode Gaussian thermal state vanishes at the same temperature regardless of whether the initial state, or the post-selected state in the discarding ensemble, is studied.

\subsection{Two oscillators: the limit of small region sizes}\label{sub:Gaussian-Small-a-Limit}

For the Gaussian system described in Section
\ref{sub:Harmonic-oscillators} the entanglement can be evaulated
analytically in the limit of very small region sizes, following the
method described in \cite{Concurrence.Density}.

\subsubsection{Only Alice's particle restricted}
Suppose only Alice makes a preliminary measurement, and determines
that her particle is located in a region of length $2a$ centred at
coordinate $\bar{q}_{A}$, as in \S\ref{sec:Method}: $\bar{q}_{A}-a\le
q_{A}\le\bar{q}_{A}+a$.  In the discarding ensemble, the entanglement
is
$E_D=h(\epsilon)\equiv-[\epsilon\log_{2}(\epsilon)+(1-\epsilon)\log_{2}(1-\epsilon)]$
with \begin{equation}
  \epsilon=a^{2}m\omega\frac{\alpha(\sqrt{1+4\alpha}-1)}{12(1+2\alpha+\sqrt{1+4\alpha})}.\label{eq:Gaussian-1}\end{equation}
Note that this depends only on $a$ and on the parameters of the
underlying oscillator system; it is independent of $\bar{q}$.  Note
also that the entanglement is non-zero for any non-zero $\alpha$, and
can be made arbitrarily large (for a given small $a$) by increasing $\alpha$.

\subsubsection{Both particles restricted}
On the other hand, if both parties make measurements, thereby also
restricting Bob's particle to a region of length $2b$ around
$\bar{q}_B$, the entanglement is once again $h(\epsilon)$ but now $\epsilon$ becomes 
\begin{equation}
  \epsilon=\frac{a^{2}b^{2}m^{2}\omega^{2}}{72}(1+2\alpha-\sqrt{1+4\alpha}),\label{eq:Gaussian-2}
\end{equation}
and the concurrence density \cite{Concurrence.Density} is
\begin{equation}
\frac{\sqrt{2}m\omega}{6}\sqrt{1+2\alpha-\sqrt{1+4\alpha}}
\end{equation}
Once again, this result depends only on the dimensionless coupling
strength $\alpha$ and the fundamental length unit $(m\omega)^{-1/2}$
of the oscillators; it is again \textit{independent} of the location
of the centres of the measurement regions. Later we will see that as
$a$ and $b$ increase, the entanglement distribution gradually changes
so that more entanglement is located at some parts of configuration
space than the others.

\subsection{Two oscillators: finite region sizes}
\subsubsection{Only Alice's particle restricted\label{sub:2HO-One-Restricted}}

For simplicity, we will set $m=1$, $\omega=1$
and choose the Gaussian characteristic length (Eq. (\ref{eq:2-14}))
for an uncoupled harmonic system, $\sigma=1$, as our unit of length.

\begin{figure*}
\includegraphics[width=0.4\linewidth,keepaspectratio]{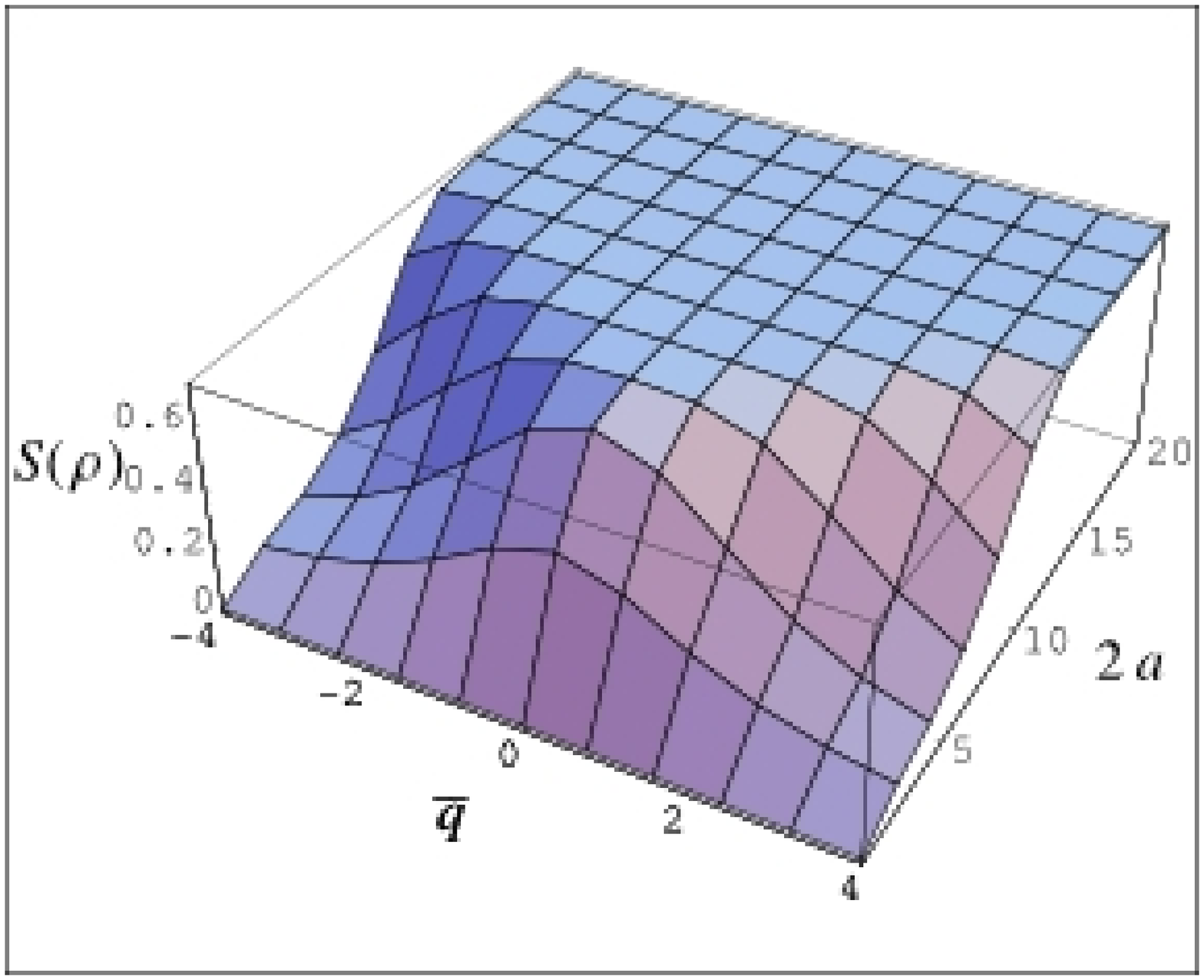}\includegraphics[width=0.4\linewidth,keepaspectratio]{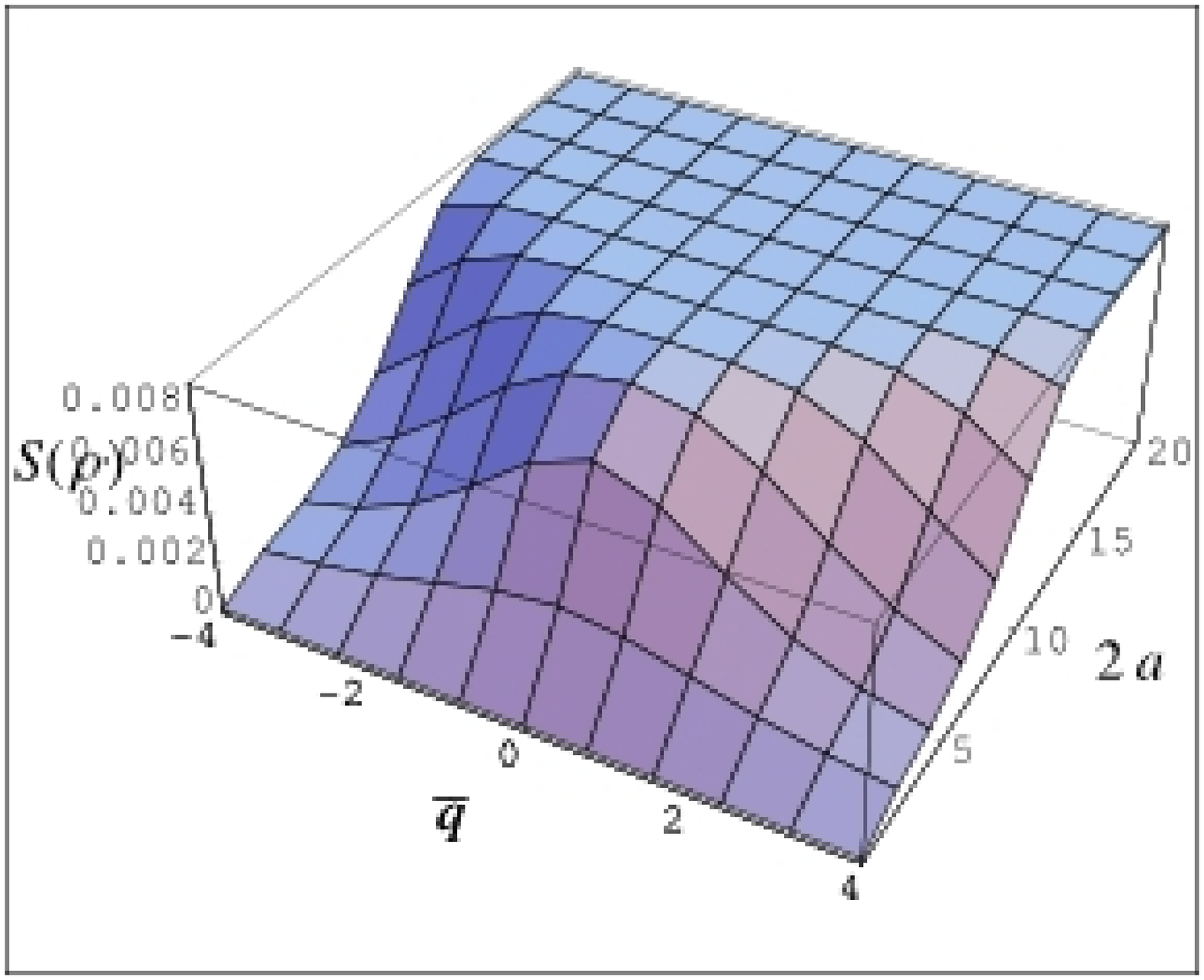}

\subfigure[$\alpha=6$]{\includegraphics[width=0.4\linewidth]{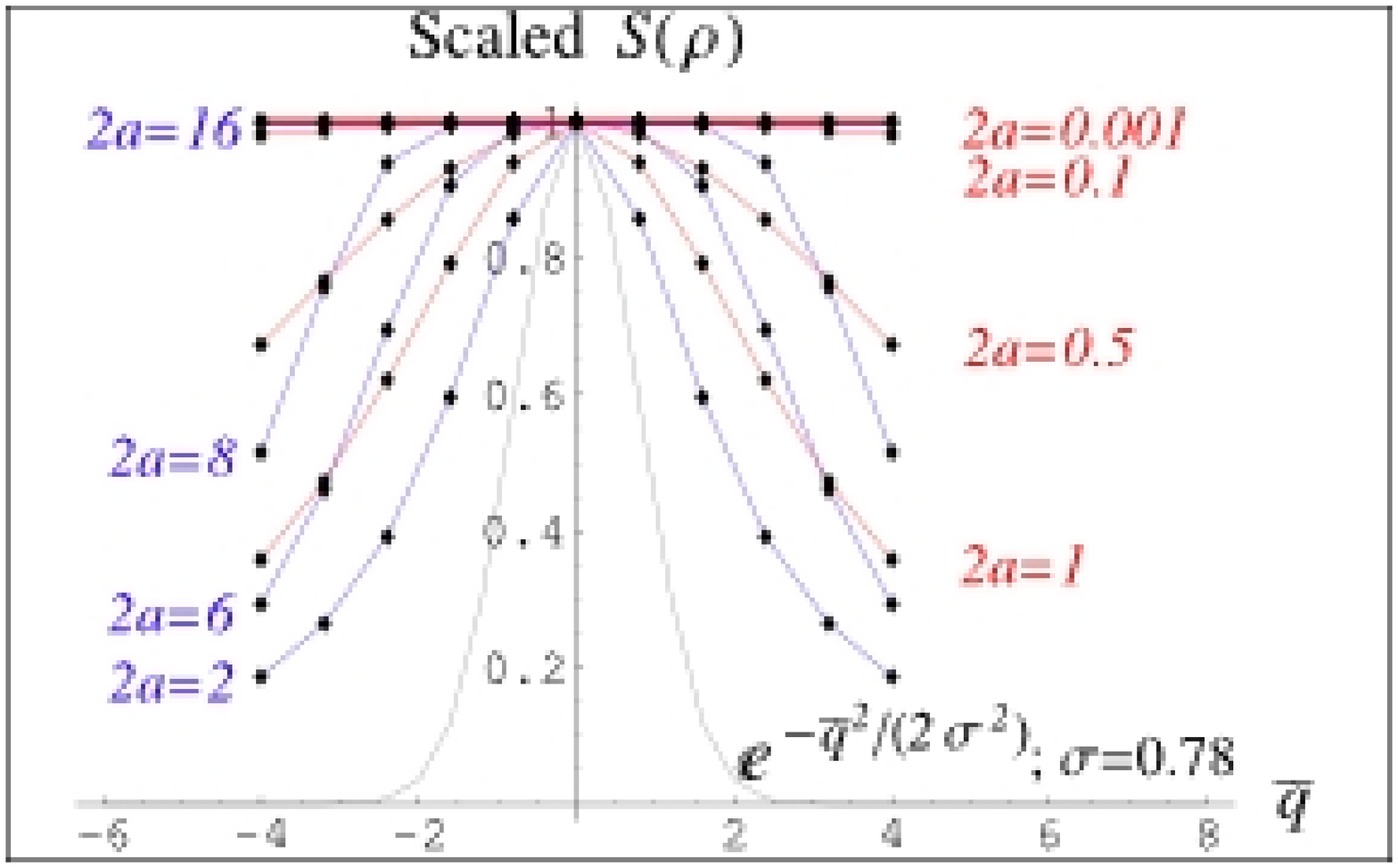}}\subfigure[$\alpha=0.06$]{\includegraphics[width=0.4\linewidth]{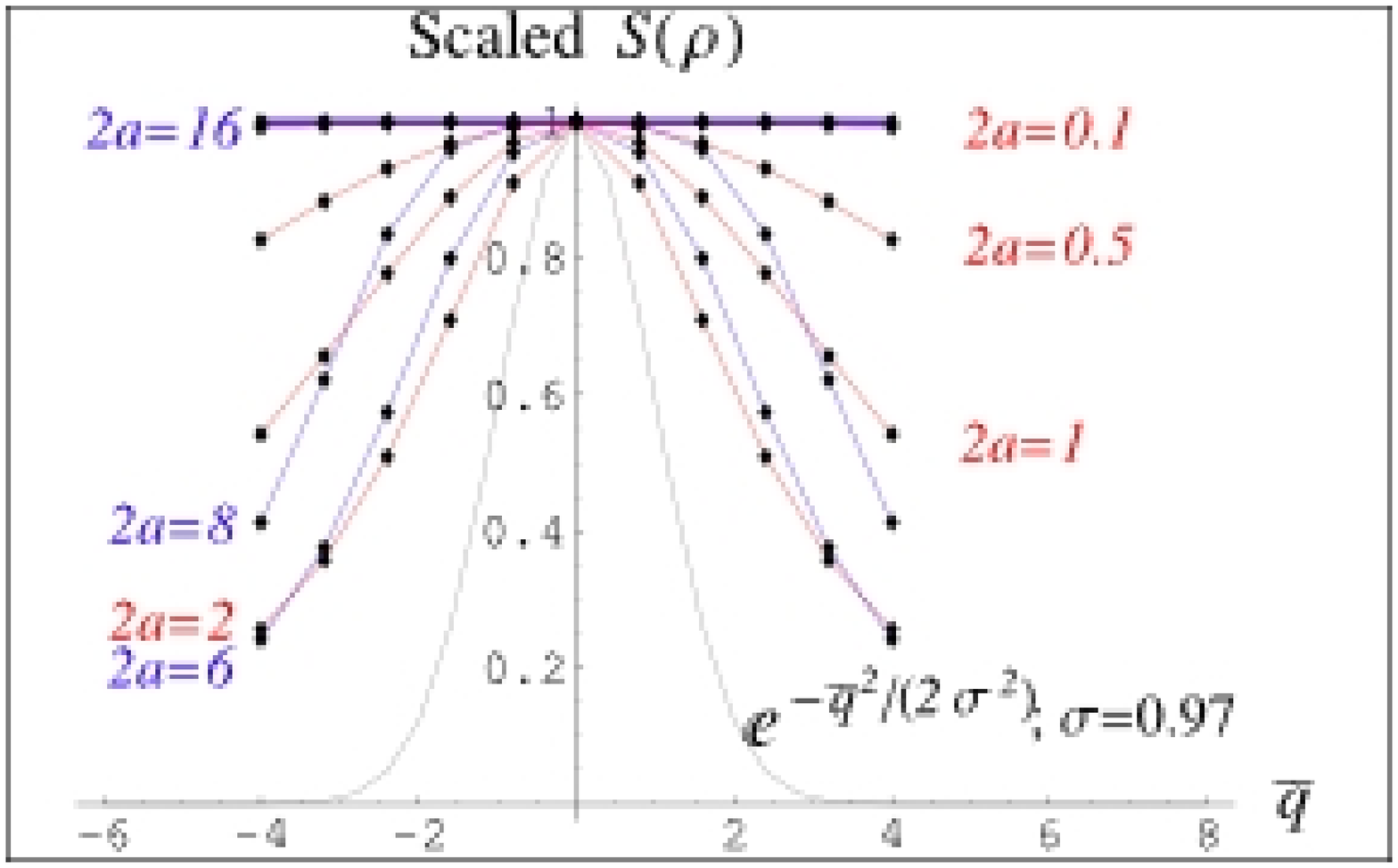}}

\caption{(Color online) Top: Variation of the entanglement $S(\rhoA)$ with both
the width $2a$ and the centre $\bar{q}$ of the preliminary-measurement region.
Bottom: $S(\rhoA)$ plotted against $\bar{q}$ for different widths,
re-scaled such that $S(\rhoA)$ has the same peak value at $\bar{q}=0$.
A plot (the black dashed line) of the corresponding Gaussian probability distribution for
Alice's particle, with a standard deviation $\sigma$ determined by
the coupling strength $\alpha$, is shown for comparison. The different
plots correspond to two different coupling strengths, (a) $\alpha=6$
v.s. (b) $\alpha=0.06$. The number of bins $N_{B}$ used in the
calculation was 200 in both cases. \label{Figure:entaglement-OneRestricted-width-position}}
\end{figure*}

In this section, we consider the case in which only Alice makes a
preliminary measurement to determine that her particle lies within a
finite-size region. Suppose that the size of this region is $2a$ and the location of
the centre of the region is $\bar{q}$, the von Neumann entropy
$S(\rhoA)$ depends on both $2a$ and $\bar{q}$. This is shown in Fig.
\ref{Figure:entaglement-OneRestricted-width-position}.  We look at the
variation with $\bar{q}$ first; Fig.
\ref{Figure:entaglement-OneRestricted-width-position} along the
$\bar{q}$-axis shows some of the examples.  For finite $a$, the
entanglement is higher if we measure around the centre of the
wavefunction, where the probability of finding a particle is
highest, than if we take our measurements further away from the the
centre of the wavefunction where the chance of finding a particle is
very low. 

We can understand this variation by examining Alice's post-selected
reduced density matrix in the centre of Fig.\
\ref{Figure:entaglement-OneRestricted-width-position} ($\bar{q}=0$)
and at the edge ($\bar{q}=\pm 4$).  At the edge, the diagonal elements
increase rapidly towards one end; the eigenvalues of this density
matrix are dominated by these terms, resulting in one eigenvalue being
close to 1 and the other eigenvalues being very small.  The von
Neumann entropy will therefore also be small.  In contrast, the
diagonal elements in the centre case, instead of being dominated by a
single element at one end, are approximately constant.  The
resulting spread of eigenvalues leads to a higher von Neumann entropy.

We would also expect that as the region size approaches the total
configuration space, the entanglement in the discarding ensemble
should tend to the entanglement originally present in the whole
system; this is shown in the upper part of Fig.
\ref{Figure:entaglement-OneRestricted-width-position}, where the
entanglement rises with $a$ until it saturates to the peak value of magnitude $S(\rhoA)=0.702$ given by Eq. (\ref{eq:gaussianeof}).  Roughly speaking, this saturation
occurs once the region has expanded to include a significant portion
of the central part of the harmonic oscillator wavefunction.

We have already seen that in the limit of small $a$ the entanglement
becomes independent of position.  In fact, even for finite $a$ the
entanglement is distributed very differently from the probability
distribution of Alice's particle. This is shown in the lower part of
Fig. \ref{Figure:entaglement-OneRestricted-width-position}, where the
coloured curves show the entanglement (scaled to a common maximum
value) as a function of $\bar{q}$ for different widths $2a$; for
comparison, the black dashed plot shows the Gaussian one-particle probability
distribution with standard deviation $\sigma$ given by Eq.
(\ref{eq:2-14}). Note that the width of the entanglement plot
varies non-monotonically with $a$: the entanglement is constant in the
limits of small and large $a$, and has a minimum width around $2a=2$ (for
$\alpha=6$).  Note also that $S(\rhoA)$ is
very small but is non-zero even for small $\alpha$, as expected from
Eq. (\ref{eq:Gaussian-1}).

For comparison, we also present in Fig.\
\ref{Figure:entaglement-OneRestricted-width-position}(b) results for a
much weaker coupling, $\alpha=0.06$ compared with $\alpha=6$: for
weak coupling, the entanglement has smaller peak values ($=0.00859$ in this case) and its
spread is narrower, but the qualitative features are similar in both
cases.

\subsubsection{Both particles restricted: entanglement distributions}

\begin{figure*}
\subfigure[$2a=0.5$]{\includegraphics[width=0.4\linewidth,keepaspectratio]{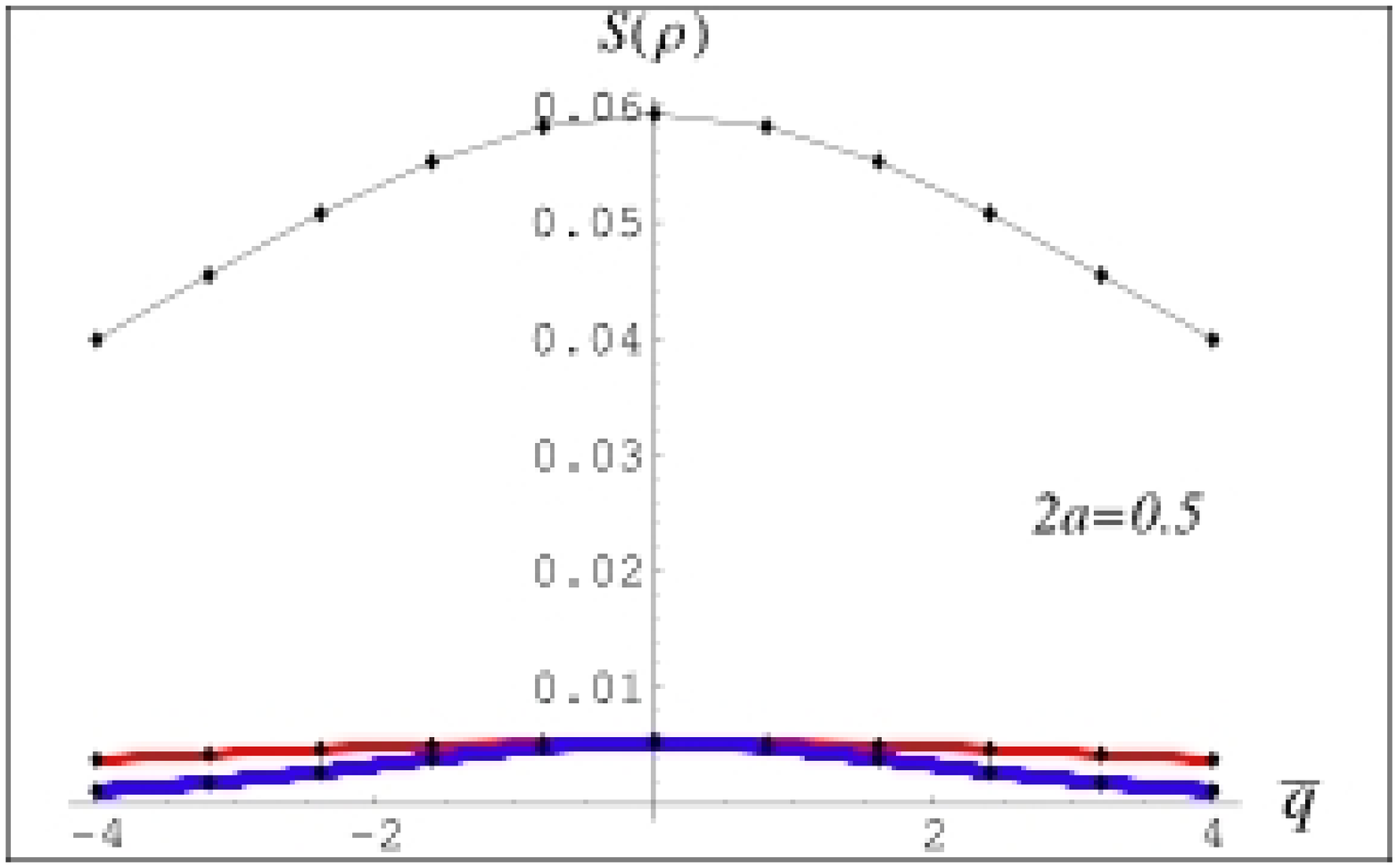}}\subfigure[Larger $2a$'s]{\includegraphics[width=0.4\linewidth,keepaspectratio]{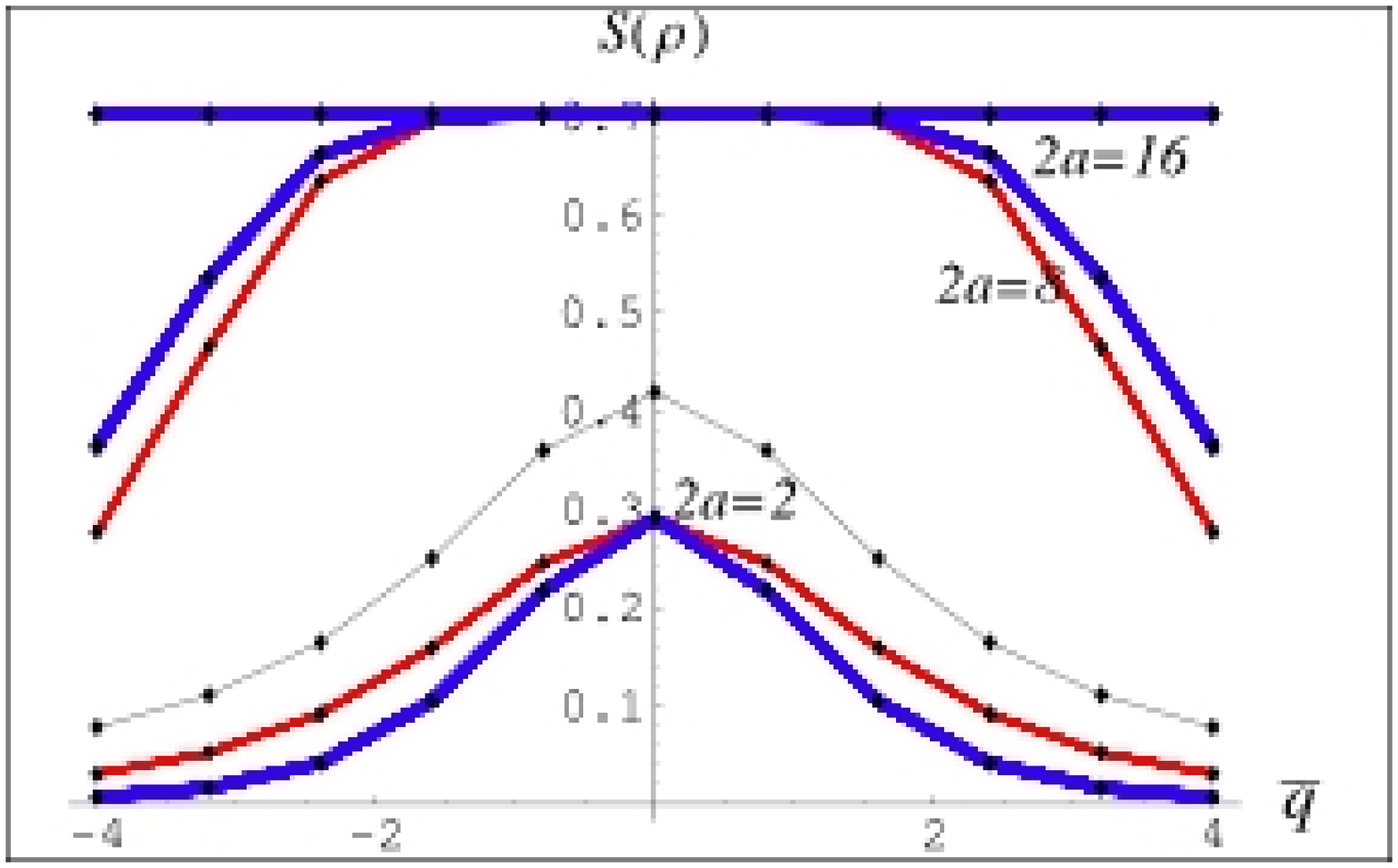}}

\caption{(Color online) Comparison of the two different cases of preliminary measurements
done by both parties together with the case that only one party makes
a preliminary measurement. The entanglement $S(\rhoA)$ is plotted
against the centre $\bar{q}$ of the preliminary-measurement region
with width $2a$. (a) For $2a=0.5$. (b) For other larger values of
$2a$. Red long-dashed line (Case 1): Both parties' preliminary measurements
localise their particles in regions with identical widths and centres
($a=b$ and $\bar{q}_{A}=\bar{q}_{B}$). Blue thick short-dashed line
(Case 2): The widths of the regions are the same but one centre is
always fixed around the centre of the wavefunction while there is
no restriction on the other centre ($a=b$, $\bar{q}_{B}=0$). Black
thin solid line (Case 3): Only one party makes a preliminary measurement.
In all three cases, the number of bins used in the calculation is
$N_{B}=100$ and $\alpha=6$.\label{Figure:entanglement-BothRestricted-Comparison-OneRestricted}}
\end{figure*}

Next we consider the case where both Alice and Bob make preliminary measurements,
but not necessarily in the same way.

We start by considering two different cases; the first (Case 1) is
that both parties' preliminary measurements restrict their particles
to regions with identical widths and centres ($a=b$ and
$\bar{q}_A=\bar{q}_B$), whereas in the second case (Case 2) the region
widths are the same but the centre of Bob's region is always fixed around the
centre of the wavefunction ($a=b$, $\bar{q}_B=0$). The
results, for $\alpha=6$, are shown together with the previous result
(Case 3; only Alice makes a preliminary measurement, as shown in Fig.
\ref{Figure:entaglement-OneRestricted-width-position}(a)) for
comparison in Fig.
\ref{Figure:entanglement-BothRestricted-Comparison-OneRestricted}.
The entanglement in the discarding ensemble of Case 3 is the highest
out of the three cases; this is as expected, since the entanglement
can only reduce under the additional (local) measurements made by Bob. When the
width $2a$ is small, the entanglement of Case 1 is higher than of Case 2.
However, as $2a$ increases, Case 2 converges more rapidly to Case 3 so that its
entanglement is now higher than that of Case 1, until $2a$ becomes so large
that the differences between all three cases disappear.

\subsubsection{Both particles restricted: classical correlations}

We now compare the entanglement distributions to the classical correlations between the particles.  Suppose that Alice and Bob
localise their respective particles to regions with the same widths
but different centres; the entanglement in the discarding ensemble
will depend on both $\bar{q}_{A}$ and
$\bar{q}_{B}$. We shall compare the entanglement distribution with the
2-particle probability distribution $P(q_{A}\in \mathcal{A}\cap q_{B}\in \mathcal{B})$,
and the conditional probability distribution for Bob's particle given
the position of Alice's particle, $P(q_{B}\in\mathcal{B}\mid q_{A}\in\mathcal{A})$.

The two-particle probability is
\begin{eqnarray}
P(q_{A}\in\mathcal{A}\cap q_{B}\in\mathcal{B}) & = & \int_{\bar{q}_{A}-a}^{\bar{q}_{A}+a}\d q_{A}\int_{\bar{q}_{B}-a}^{\bar{q}_{B}+a}\d q_{B}\nonumber \\
&&\quad\rho(q_{A},q_{B};q_{A},q_{B}),\label{eq:BPMR-1}\end{eqnarray}
 and in the limit of small $a,b$ we have 
\begin{equation}
P(q_{A}\in \mathcal{A}\cap q_{B}\in \mathcal{B})=4ab\rho(\bar{q}_A,\bar{q}_B;\bar{q}_A,\bar{q}_B).
\end{equation}

The conditional probability is
\begin{equation}
P(q_{B}\in\mathcal{B}\mid q_{A}\in\mathcal{A})=\frac{P(q_{A}\in \mathcal{A}\cap q_{B}\in \mathcal{B})}{P(q_{A}\in\mathcal{A})},\label{eq:BPMR-2}\end{equation}
where $P(q_{A}\in\mathcal{A})$ is the 1-particle probability.  In the limit of small $a,b$ this becomes
\begin{equation}
P(q_{B}\in\mathcal{B}\mid q_{A}\in\mathcal{A})=2b\frac{\rho(\bar{q}_A,\bar{q}_B;\bar{q}_A,\bar{q}_B)}{\rho^{(A)}(\bar{q}_A,\bar{q}_A)}.
\end{equation}
In each case the small-$a,b$ limit can be easily evaluated: we find
\begin{equation}
\rho(q_{A},q_{B};q_{A},q_{B})=\zeta_{2}\exp(-\frac{(\bar{q}_{A}+\bar{q}_{B})^{2}}{2\sigma_{+}^{2}}-\frac{(\bar{q}_{A}-\bar{q}_{B})^{2}}{2\sigma_{-}^{2}})\label{eq:BPMR-3}\end{equation}
with `classical' standard deviations
\begin{eqnarray}
\sigma_{+}^{C} & = & \sqrt{2},\nonumber \\
\sigma_{-}^{C} & = &
(\frac{2}{\sqrt{1+4\alpha}})^{\frac{1}{2}},\label{eq:fitC}
\end{eqnarray}
and
\begin{equation}
\frac{\rho(\bar{q}_A,\bar{q}_B;\bar{q}_A,\bar{q}_B)}{\rho^{(A)}(\bar{q}_A,\bar{q}_A)}=\zeta_{3}\exp(-\frac{\bar{q}_{A}^{2}}{2\sigma_{1}^{2}}+\frac{\bar{q}_{A}\bar{q}_{B}}{2\sigma_{12}^{2}}-\frac{\bar{q}_{B}^{2}}{2\sigma_{2}^{2}})\label{eq:fit2}
\end{equation}
with
\begin{eqnarray}
\sigma_{1} & = & (\frac{(1+\sqrt{1+4\alpha})(1+2\alpha+\sqrt{1+4\alpha})}{4\alpha^{2}})^{\frac{1}{2}},\label{eq:BPMR-5}\\
\sigma_{2} & = & (\frac{2}{1+\sqrt{1+4\alpha}})^{\frac{1}{2}},\nonumber \\
\sigma_{12} & = & (\frac{1}{\sqrt{1+4\alpha}-1})^{\frac{1}{2}},\nonumber \end{eqnarray}
where $\zeta_2$ and $\zeta_3$ are normalization constants.  

For finite $a$ and $b$ we capture the shape of the distributions by
fitting the numerically calculated values of $P(q_{A}\in
\mathcal{A}\cap q_{B}\in \mathcal{B})$, and $P(q_{B}\in\mathcal{B}\mid
q_{A}\in\mathcal{A})$ using the same expressions Eq. (\ref{eq:BPMR-3}) and Eq. (\ref{eq:fit2}), thereby extracting numerical values for
$\sigma^C_{\pm}$, $\sigma_{1,2}$ and $\sigma_{12}$.  We also use the
function Eq. (\ref{eq:BPMR-3}) to fit the entanglement distribution,
thereby obtaining two further parameters $\sigma^{Q}_\pm$ which
quantify the extent of the entanglement distribution along its
principal axes.

\begin{figure}[ht]
\subfigure[$2a=0.5$]{\includegraphics[width=0.5\linewidth,keepaspectratio]{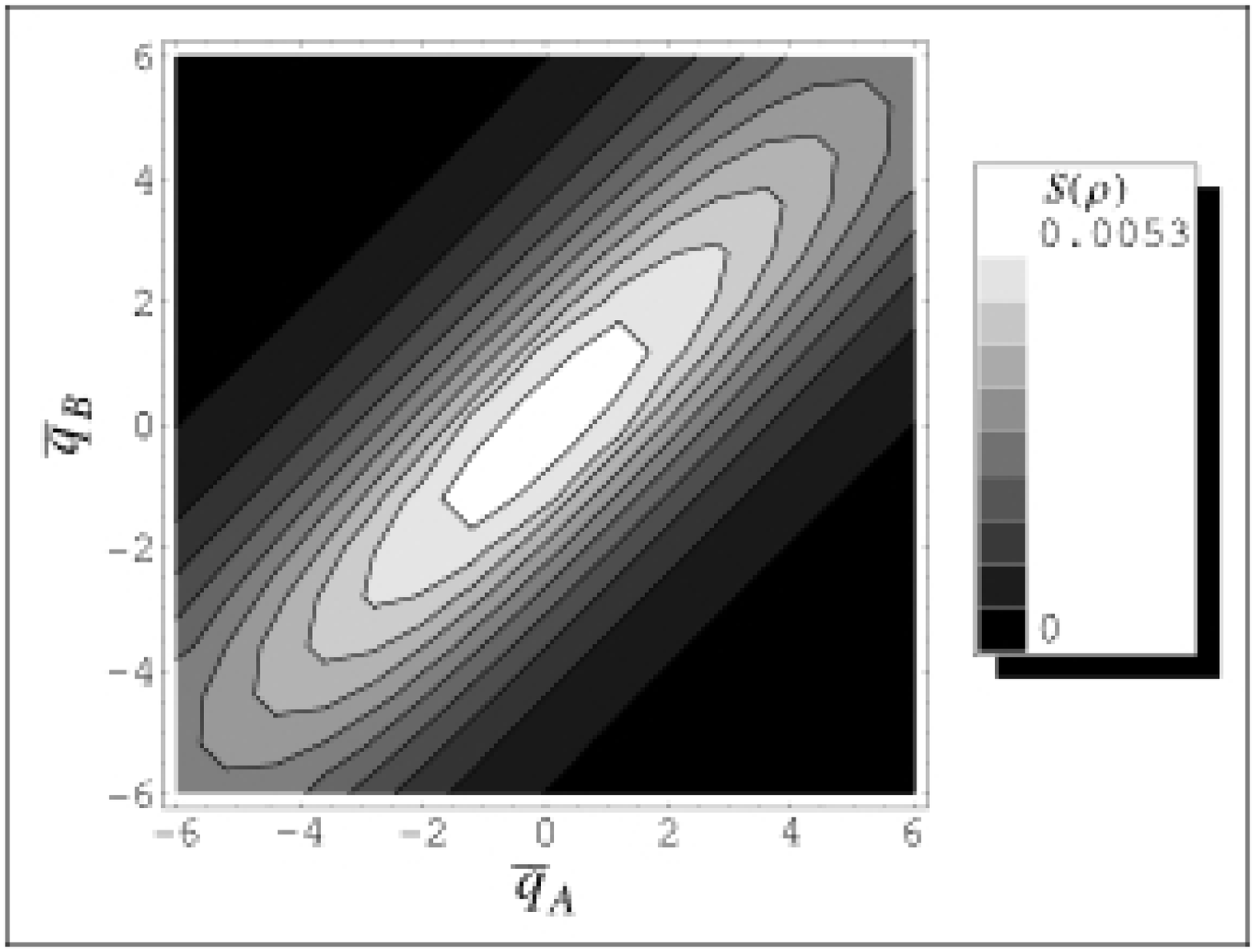}}\subfigure[$2a=4$]{\includegraphics[width=0.5\linewidth,keepaspectratio]{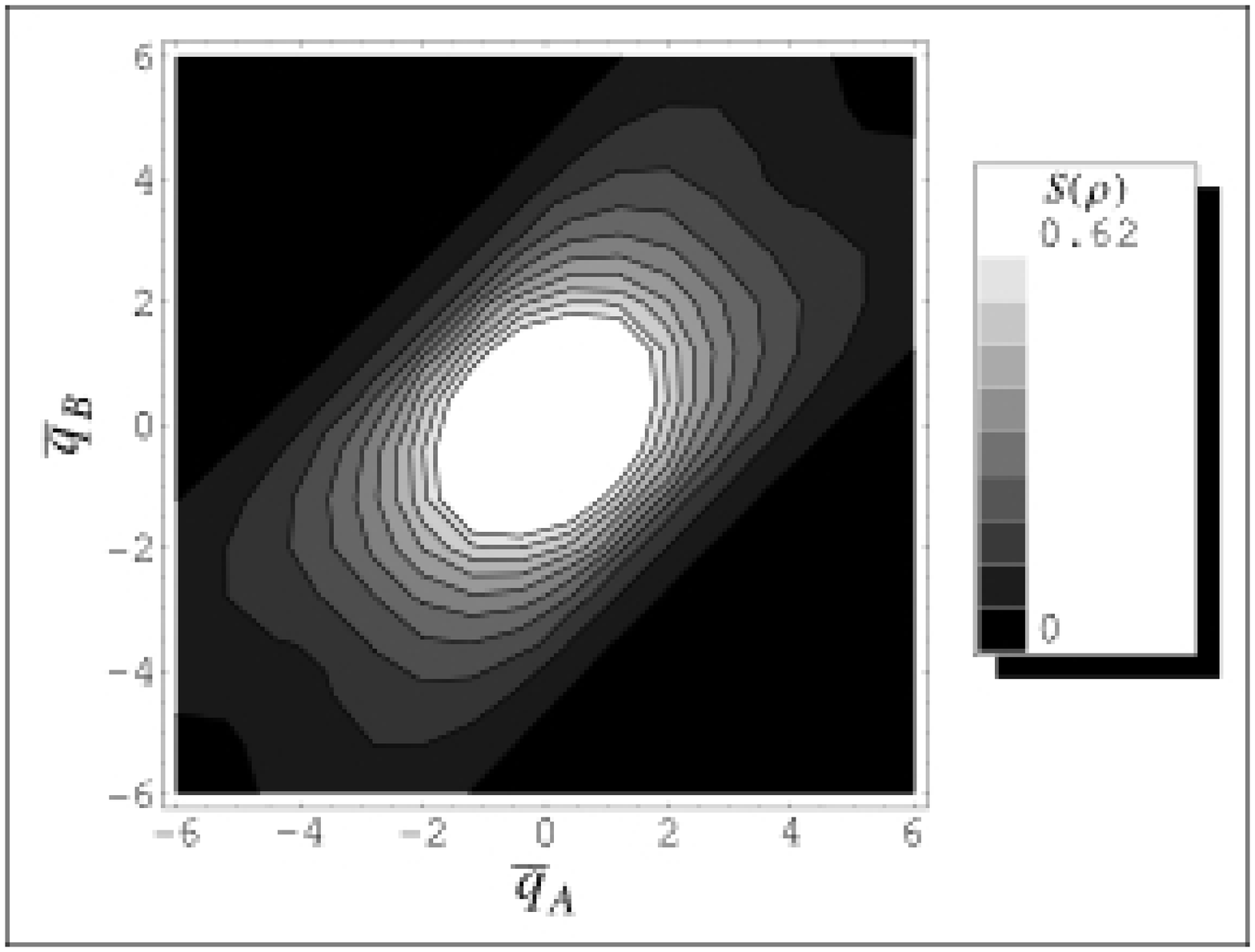}}

\caption{Dependence of the entanglement $S(\rhoA)$ on the locations of the centres of
  the preliminary-measurement regions $\bar{q}_{A}$ and $\bar{q}_{B}$.
  (a) the width $2a$ of the regions is 0.5. (b) $2a=4$. In both cases,
  $N_{B}=100$ and $\alpha=6$.\label{Figure:Quantum-Correlations}}
\end{figure}

\begin{figure}[ht]
\subfigure[$2a=0.5$]{\includegraphics[width=0.5\linewidth,keepaspectratio]{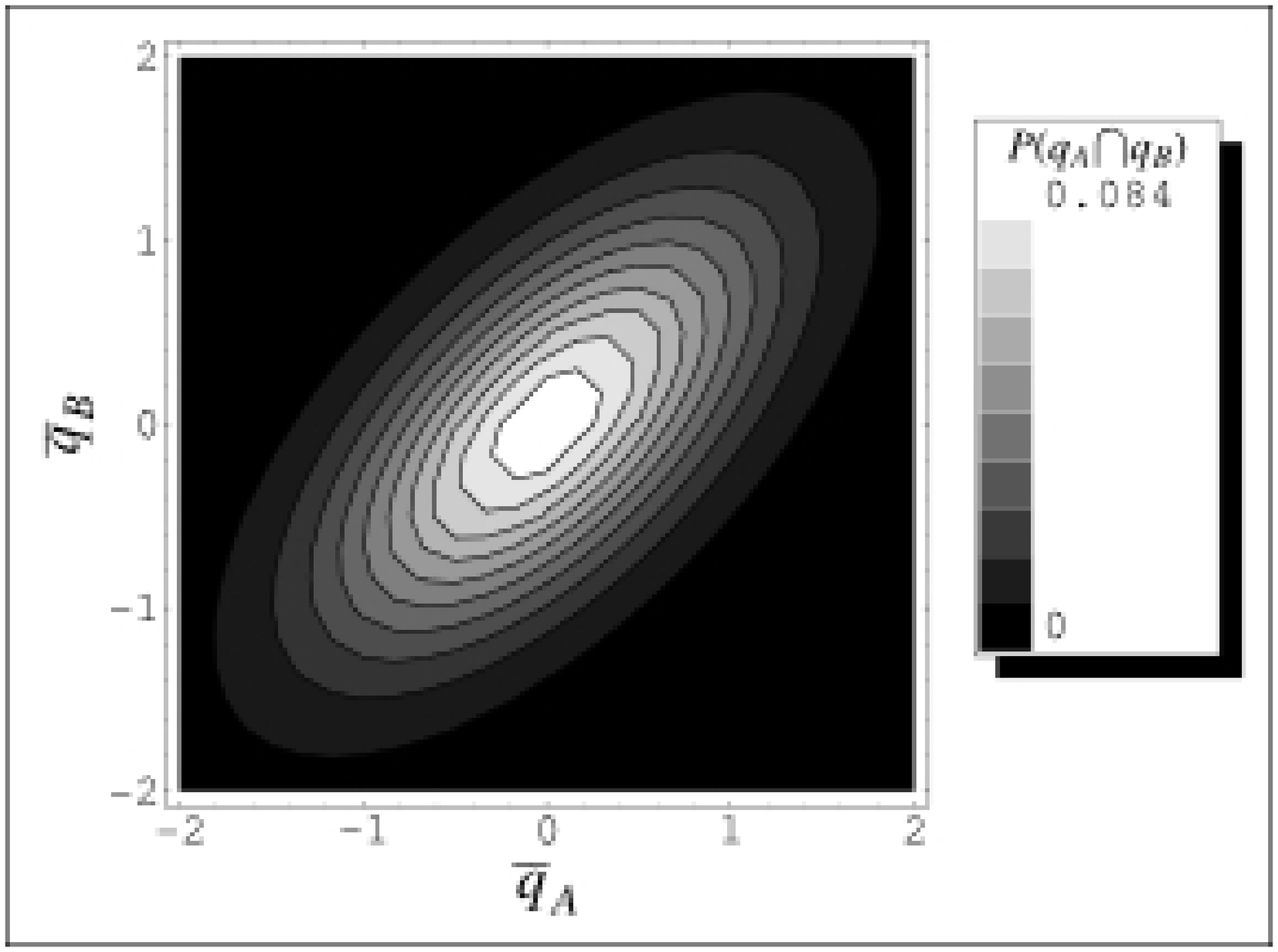}}\subfigure[$2a=4$]{\includegraphics[width=0.5\linewidth,keepaspectratio]{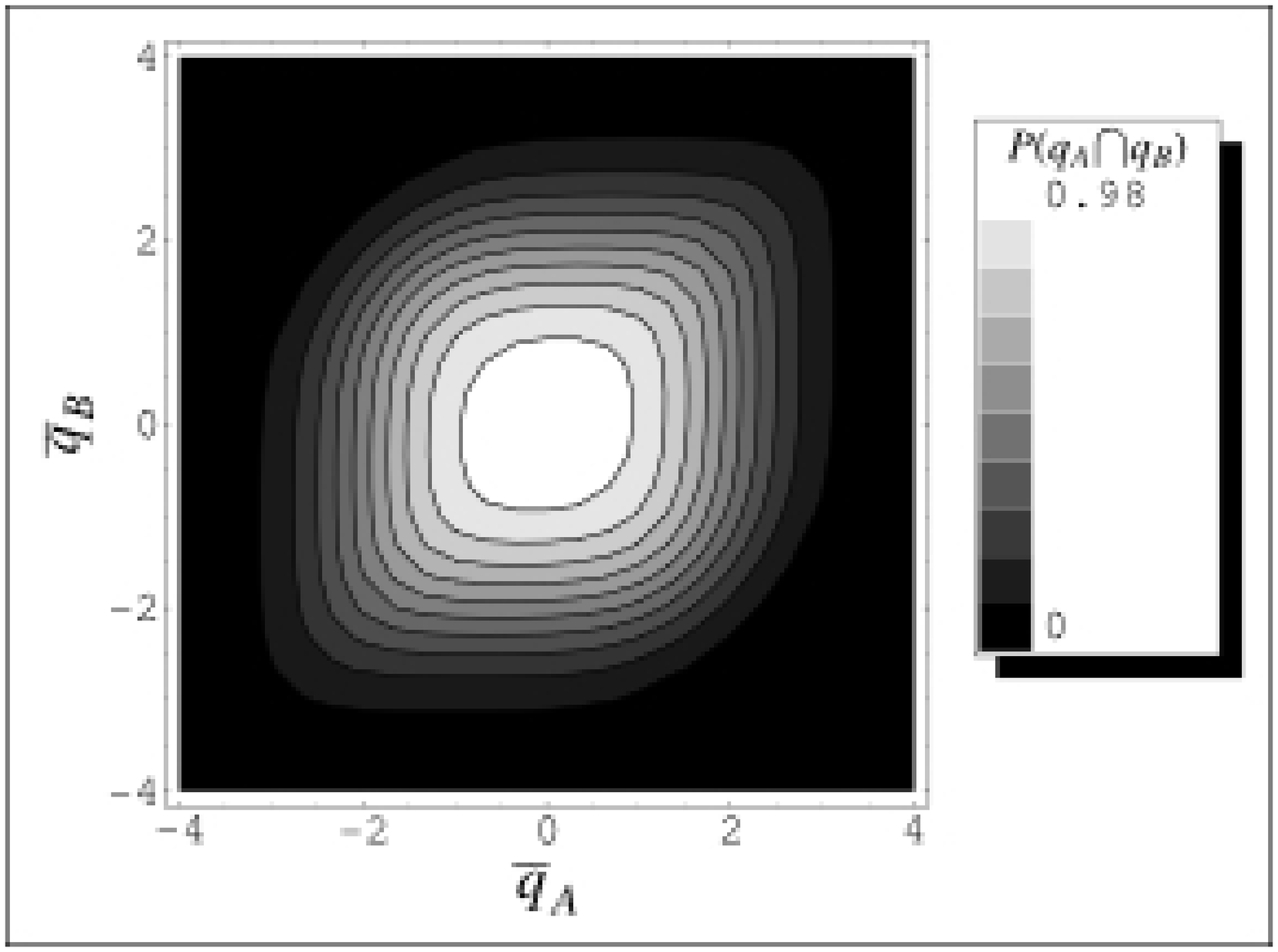}}

\caption{The dependence of the classical joint probability
  $P(q_{A}\in\mathcal{A}\cap q_{B}\in\mathcal{B})$
on $\bar{q}_{A}$ and $\bar{q}_{B}$. (a) $2a=0.5$. (b) $2a=4$. In both cases,
$\alpha=6$.\label{Figure:Classical-Correlations}}
\end{figure}

\begin{figure}[ht]
\subfigure[$2a=0.5$]{\includegraphics[width=0.5\linewidth,keepaspectratio]{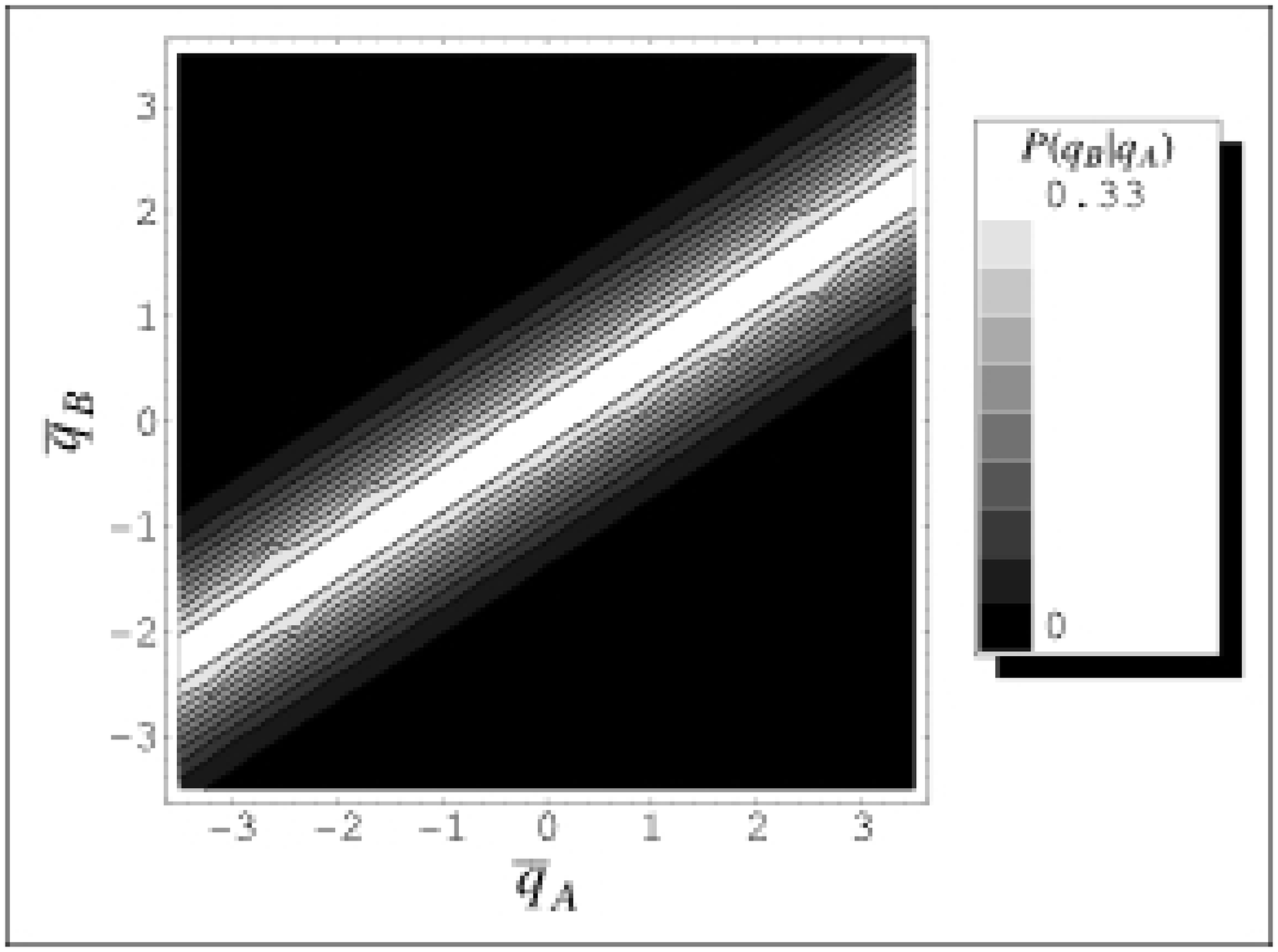}}\subfigure[$2a=4$]{\includegraphics[width=0.5\linewidth,keepaspectratio]{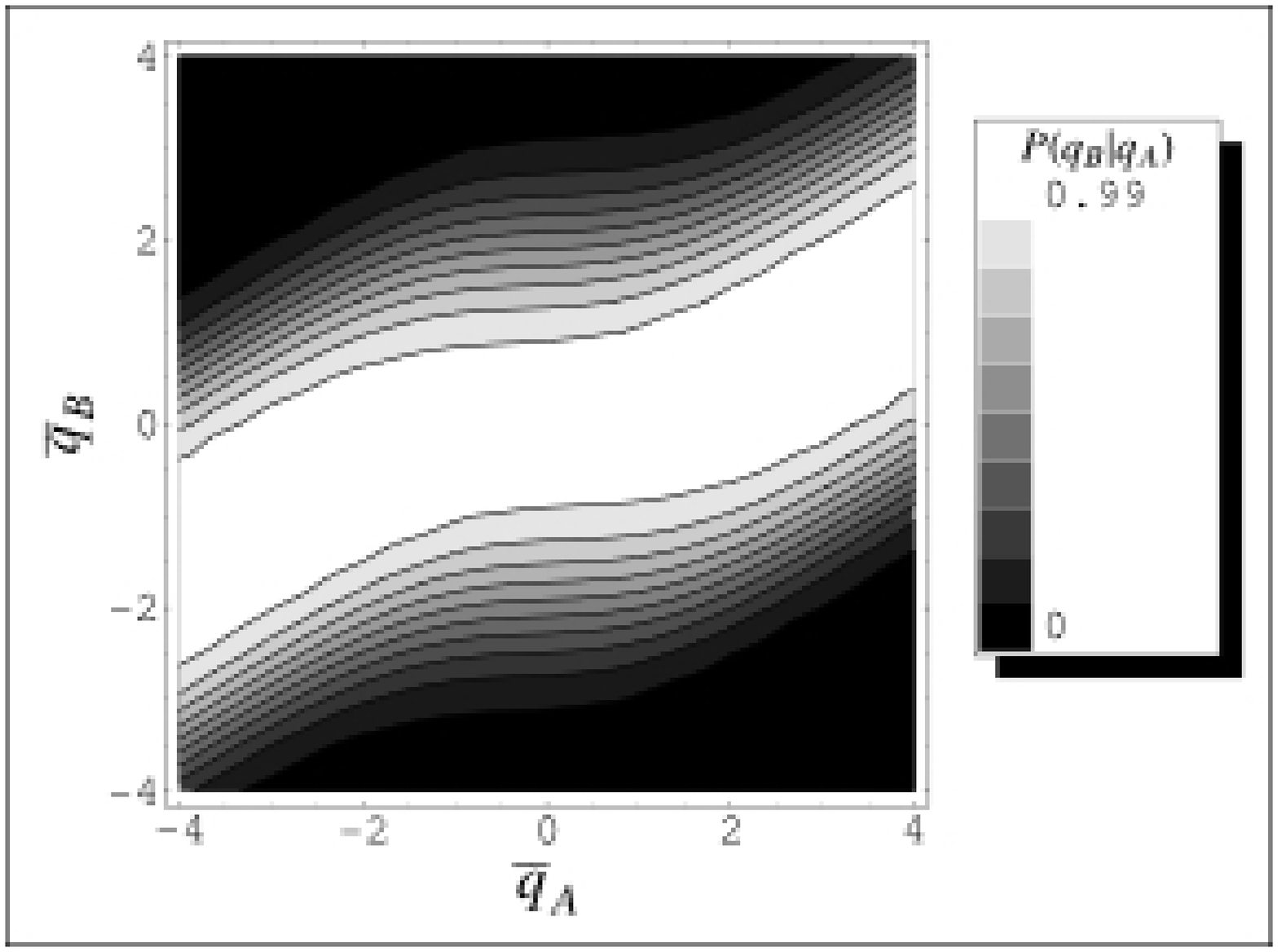}}

\caption{Dependence of the conditional probability
  $P(q_{B}\in\mathcal{B}\mid q_{A}\in\mathcal{A})$ on $\bar{q}_{A}$
  and $\bar{q}_{B}$. (a) $2a=0.5$. (b) $2a=4$. In both cases,
  $\alpha=6$.\label{Figure:Classical-Correlations-Conditional}}
\end{figure}

As before, we take $\alpha=6$. In Fig.
\ref{Figure:Quantum-Correlations}, we show two cases of entanglement
distributions for different widths ($2a=0.5$ and $2a=4$) of the
preliminary-measurement regions. We see that the entanglement
distribution with larger $2a$ is more symmetric.  The corresponding joint probability distributions and conditional probability distributions
are shown respectively in Fig. \ref{Figure:Classical-Correlations} and
Fig. \ref{Figure:Classical-Correlations-Conditional}. (Note that the figures show different range of $\bar{q}_{A}$ and $\bar{q}_{B}$.) The classical probability
distributions $P(\bar{q}_{A}\cap\bar{q}_{B})$ are more localized and
symmetric in space than the entanglement distributions.

\begin{table}[ht]
\begin{tabular}{|c|c|c|c|c|c|c|c|}
\hline 
$\alpha=6$&
$\sigma_{+}^{Q}$&
$\sigma_{-}^{Q}$&
$\sigma_{+}^{C}$&
$\sigma_{-}^{C}$&
$\sigma_{1}$&
$\sigma_{2}$&
$\sigma_{12}$\tabularnewline
\hline
\hline 
$2a\rightarrow0$&
$\infty$&
$\infty$&
1.41&
0.632&
0.866&
0.577&
0.500\tabularnewline
\hline 
$2a=0.5$&
10.4&
2.29&
1.43&
0.665&
0.937&
0.603&
0.531\tabularnewline
\hline 
$2a=4$&
3.44&
2.10&
2.37&
2.00&
11.0&
1.53&
2.64\tabularnewline
\hline
\end{tabular}

\caption{Table of $\sigma$ values for $\alpha=6$.\label{table:sigmavals}}
\end{table}

In the limit of very small $a$, $S(\rhoA)$ is constant everywhere (Eq.
(\ref{eq:Gaussian-2})) so $\sigma^Q_{+}$ and $\sigma^Q_{-}$ must
diverge; the results in Table\ \ref{table:sigmavals} show that
$\sigma^Q_+$ diverges more quickly as $a$ reduces, while the two
parameters become comparable for large $a$ as the entanglement
distribution becomes more symmetric.  Indeed, the distributions of the
entanglement and the classical correlations become more alike as $2a$
increases, because both distributions are flat out to a distance $a$
either side of the wavefunction's central peak.

\begin{figure}[ht]
\subfigure[$\sigma^Q_{\pm}$; $2a=0.5$]{\includegraphics[width=0.8\linewidth,keepaspectratio]{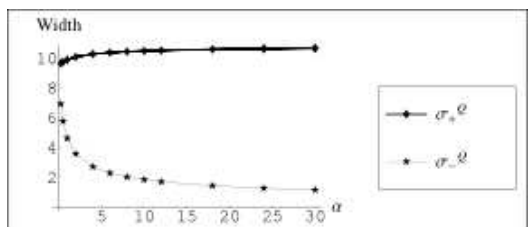}}

\subfigure[$2a=0.5$]{\includegraphics[width=0.8\linewidth,keepaspectratio]{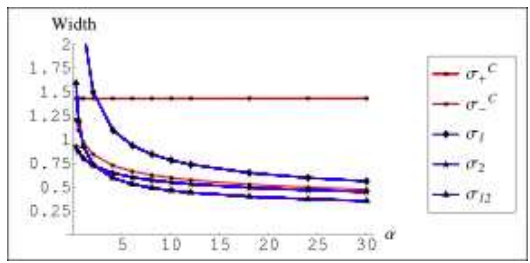}}

\subfigure[Small-$a$ limit]{\includegraphics[width=0.8\linewidth,keepaspectratio]{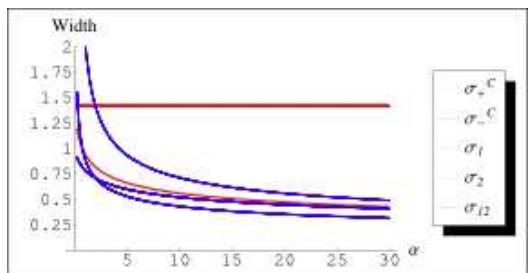}}

\caption{(Color online) Plots of $\sigma_{+}$, $\sigma_{-}$, $\sigma_{1}$, $\sigma_{12}$
and $\sigma_{2}$ against $\alpha$. In the plot legend, $Q$ stands
for the `quantum' entanglement distribution and $C$ for the `classical' probability distribution. (a) and
(b): Numerical results: $2a$ is chosen to be 0.5 for all the cases.
(c) Analytical results: in the limit of very small $a$. \label{Figure:Sigmas-Alpha-Quantum-Classical}}
\end{figure}

We can also study the effect of varying 
the coupling strength $\alpha$ for a fixed (small) $2a$.  We plot $\sigma_{+}^{Q}$ and $\sigma_{-}^{Q}$ against $\alpha$ with $2a=0.5$ in Fig. \ref{Figure:Sigmas-Alpha-Quantum-Classical}a whereas $\sigma_{+}^{C}$, $\sigma_{-}^{C}$,  $\sigma_{1}$, $\sigma_{2}$
and $\sigma_{12}$ in Fig. \ref{Figure:Sigmas-Alpha-Quantum-Classical}b.
The entanglement distribution is the most asymmetrical and as $\alpha$
increases, the difference between $\sigma_{+}^{Q}$ and $\sigma_{-}^{Q}$
widens. Of the quantities determining the classical probability distribution, $\sigma_{+}^{C}$
remains constant with increasing $\alpha$, but $\sigma_{-}^{C}$ gradually
decreases. These trends arise because the two particles
tend to move together when the spring joining them becomes strong.
Therefore, as $\alpha$ increases, the white rod
in Fig. \ref{Figure:Classical-Correlations-Conditional} rotates about
the centre of the square from the line $\bar{q}_{B}=0$ towards the diagonal $\bar{q}_{A}=\bar{q}_{B}$. $\sigma_{1}$ is always the largest out
of the three parameters for the conditional probability distribution.
For weak $\alpha$, $\sigma_{12}$ is larger than $\sigma_{2}$ but
as $\alpha$ becomes larger, at some point the two plots intercept
and $\sigma_{12}$ is no longer larger than $\sigma_{2}$. 

How in the limit of very small $a$ these quantities (Eq. (\ref{eq:fitC}) and Eq. (\ref{eq:BPMR-5})) vary with $\alpha$ is shown in Fig. \ref{Figure:Sigmas-Alpha-Quantum-Classical}c.
We see that the behaviour of these quantities do not change much,
compared with the previous results when $2a=0.5$, apart from that
the interception points happen at smaller $\alpha$. Note that $\sigma^Q_\pm$ diverge as $a\rightarrow 0$, so these parameters are not shown.

\section{Conclusions}

We have presented a novel thought experiment that gives an approach
to determining the location in configuration space of the entanglement
between two systems. It involves choosing a region of the two-party
configuration space and making a projective measurement with only
enough resolution to determine whether or not the system resides in
this region, then characterizing the entanglement remaining in the
corresponding sub-ensemble. Our approach is particularly simple to
implement for pure states, since in this case the sub-ensemble in
which the system is definitely located in the required region after
the measurement is also a pure state, and hence its entanglement can
be simply characterized by the entropy of the reduced density operators.

We have given examples of the application of our method to states
of a simple spin system, where Alice and Bob share two pairs of spin-1/2
particles, and also of a continuous-variable system in which they
share a pair of coupled harmonic oscillators. 

The first case shows
how the amount of entanglement located in the chosen region (in this
case the $M_{s}=0$ manifold) varies as the characteristics of the
states shared by Alice and Bob are altered. We presented results for both pure and mixed states, and show how entanglement is affected by the ``mixedness'' $F$ both qualitativly and quantitativly. Specifically, we  show that the states which are entangled from the global point of view are also entangled by our local measures \cite{Concurrence.Density}, i.e. global entanglement of the initial state vanishes at the same point as the entanglement remaining in the discarding ensemble after the preliminary measurement to locate the system in a chosen subspace. 

For the second case we have presented results as a function
of the strength of the coupling between the oscillators, as well as
of the size and location of the preliminary measurement regions. In
all cases the remaining entanglement saturates to the total entanglement
of the system as the measured regions become large. For small measured
regions the entanglement tends to zero, but for a fixed region size
the configuration-space location can be varied in order to give a
variable-resolution map of the entanglement distribution. We find that
the distribution of the entanglement is qualitatively different from
the classical correlations between the particles, being considerably
more extended in configuration space than the joint probability density and becoming more and more diffuse as the size of the regions decreases.

Our approach suffers from the disadvantage that there is no sum rule
on the entanglements in the discarding ensemble: the sum of the entanglements from all the sub-regions defined by a given decomposition
of configuration space does not yield the full entanglement
of the system. Instead, the entanglements from the sub-regions satisfy the inequality in Eq.\ (\ref{eq:discardinginequality}). It would be interesting to understand in more detail the relationship between the restricted entanglements (as defined in this paper) and the full entanglement of the system, and also to extend the calculations reported here
to projective measurements made in other bases, to POVMs, and to mixed
states. 

\begin{acknowledgments}
We are grateful to Sougato Bose for a number of valuable discussions.
\end{acknowledgments}
\bibliographystyle{apsrev}
\bibliography{2HO_entanglement_restricted_measurements_paper}
%\bibite

\end{document}